\documentclass[usenatbib]{mnras}

\usepackage{amsmath}
\usepackage{graphicx}

\usepackage{mathptmx}
\usepackage{txfonts}

\usepackage[T1]{fontenc}
\usepackage{ae,aecompl}

\usepackage{color}

\newcommand{\edit}[1]{#1}

\newcommand{\eq}[1]{(\ref{#1})}

\defcitealias{Pil2016}{EPL16}

\newcommand{\emax}{e_{\mathrm{max}}}

\newcommand{\mps}{m\,s$^{-1}$}

\title[Secular resonances in binary star systems]{%
Dynamics and habitability in circumstellar planetary systems of known binary stars
}

\author[{\'A}. Bazs{\'o} et al.]{%
{\'A}kos Bazs{\'o},$^{1}$\thanks{E-mail: bazso@astro.univie.ac.at}
Elke Pilat-Lohinger,$^{1}$
Siegfried Eggl,$^{2}$
\newauthor
Barbara Funk,$^{1}$
David Bancelin,$^{1}$
and Gioia Rau$^{1}$
\\
$^{1}$Institute for Astrophysics, University of Vienna, T{\"u}rkenschanzstra{\ss}e 17, A-1180 Vienna, Austria\\
$^{2}$IMCCE, Observatoire de Paris, 77 Avenue Denfert-Rochereau, F-75014 Paris, France
}

\date{Accepted XXX. Received YYY; in original form ZZZ}

\pubyear{2016}

\begin{document}
\label{firstpage}
\pagerange{\pageref{firstpage}--\pageref{lastpage}}
\maketitle

% -----------------------------------------------------------------------------

\begin{abstract}
We present a survey on exoplanetary systems of binary stars with stellar separations less than 100~au. For a sample of 11 binaries that harbour detected circumstellar giant planets we investigate the frequency of systems with secular resonances (SR) affecting the habitable zone (HZ). Secular resonances are connected to dynamically unstable or chaotic regions by enforcing highly eccentric motion. We apply a semi-analytical method to determine the locations of linear SR, which is based on finding the apsidal precession frequencies of the massive bodies. For configurations where the giant planet is located exterior to the HZ we find that there is always a SR interior to its orbit, the exact location of the SR strongly depends on the system's architecture. In systems with the giant planet interior to the HZ no SR can occur in the Newtonian framework. Taking into account the general relativistic precession of the perihelion, which increases the precession frequencies, planets with $a < 0.1$~au can cause SR in the HZ. We find two cases where the SR is located inside the HZ, and some more where it is close to the HZ. Generally, giant planets interior to the HZ are more favourable than exterior planets to avoid SR in the HZ. Around the location of the SR weaker mean-motion resonances are excited, and resonance overlap is possible. Existing analytical models are not as accurate as the semi-analytical method in locating the SR and deviate by $\sim 0.1$~au or more.
\end{abstract}

\begin{keywords}
binaries: general -- celestial mechanics -- planets and satellites: dynamical evolution and stability 
\end{keywords}
% stars: individual by HD number
% {HD~1257, HD~13445, HD~19994, HD~41004, HD~120136, HD~126614, HD~128621, HD~177830, HD~196885, HD~222404, Kepler-420}

% -----------------------------------------------------------------------------

% ----- BODY OF PAPER -----

% section "Introduction"
% -----------------------------------------------------------------------------
% FILE    : intro.tex
% PURPOSE : binary star article, section "Introduction"
% DATE    : 22 Nov 2016
% -----------------------------------------------------------------------------

\section{Introduction}\label{sec:intro}

% --- exoplanet diversity ---
Extrasolar planets (exoplanets) orbiting other stars exist in a large variety of different configurations. Some exoplanets orbit their host star with periods of only a few days, e.g. the Hot Jupiters \citep[see][]{Bea2012b, Naoz2012}. Other exoplanets were observed on wide orbits at large distances from their host star \citep[e.g. HR~8799,][]{Mar2008}; however, their formation mechanism is still poorly understood \citep{Dod2009, Vor2013}.

% -------------------------------------

% --- multiplicity of exoplanet host stars ---
An important point is that exoplanets are not exclusive companions to single stars, but also binary and multiple stars host planetary objects \citep{Des2007, Roe2012}. A considerable fraction of stars in the solar neighbourhood are members of binary and multiple star systems. Observational surveys derived a percentage of multiple stars of about 45~\% \citep{Duq1991, Rag2010, Duc2013, Tok2014}. Of course, not all of those systems are hosting planets. Recent estimates on the multiplicity of exoplanet host stars yield the percentage of double stars to be 10--15~\%, while multiple systems make up for about 2~\% \citep{Rag2006, Mug2009, Roe2012}.

% -------------------------------------

% --- planet formation and binary stars ---
Although exoplanets in binary and multiple star systems have been found, it is not clear whether or not these environments are more hostile for the presence of planets than single stars \citep{Boss2006, Bro2015, Jan2015}. The planet formation process in binaries is basically proceeding in the same way as for single stars \citep{The2015}, however the presence of a companion star can truncate the proto-planetary disk \citep{Kley2010} and influence planet formation \citep{Mul2012}, possibly limiting the occurrence rate of planets \citep[see][]{Winn2015}. \citet{Wang2014} found that planets in multiple star systems with stellar separations between 10--1000 astronomical units (au) occur about a factor of two less frequently compared to single stars. On the other hand, exoplanets may still form in `tight' binaries with separations of $\lesssim 30$~au. \citet{Jan2015} calculated the likelihood that giant planets can form from the truncated disk in such tight binaries. The results indicate that the disks still contain enough material to form giant planets by core accretion, while terrestrial planets can form in even more depleted disks.

% -------------------------------------

% --- stability studies  ---
Detected exoplanets in binary star systems can be divided into three dynamical classes, of which S-type \citep{Rabl1988} and P-type \citep{Dvo1989} are the most important cases. S-type planets are circumstellar planets; they orbit one component of the binary star system, just as a planet would orbit a single star. P-type planets are circumbinary planets that orbit both stars in some distance. In both cases, long-term stable planetary orbits may occupy only certain regions. \citet{Rabl1988} established an empirical formula for the size of the circumstellar region where stable planetary motion can occur in an eccentric binary system for equal mass stars. \citet{Hol1999} generalized those results to both S-type and P-type binaries with variable mass-ratios, and gave formulas for the edge of the stable region as function of binary separation, eccentricity, and mass-ratio. \citet{Pil2002} also included the influence of a non-zero initial planetary eccentricity on the orbital stability region, which can shrink considerably for large mass-ratios and eccentric planets relative to the circular case.

% -------------------------------------

% --- secular dynamics of exoplanetary systems ---
Albeit the orbital stability of exoplanets is an intriguing topic in itself, it is not the only prerequisite for habitability. Substantial changes in insolation that are due to variations in planetary orbits have to be taken into account as well \citep{Wil2002}.
\edit{This tight coupling between orbital dynamics and habitability is demonstrated by \citet{For2016}, who uses a coupled model for spin-orbit-radiative perturbations to investigate Milankovitch cycles of P- and S-type planets in binary stars. The stellar companion induces temperature oscillations on the planet on short and long time-scales via changes in the obliquity and precession.}

\edit{Since orbital dynamics is also important to assess habitability}, we need to consider the effect of both mean-motion resonances (MMR, involving orbital frequencies) and secular resonances (SR, related to apsidal frequencies) on the planet's orbit. Secular resonances act on time-scales comparatively long to a planet's orbital period. In connection with dynamically unstable or chaotic regions in planetary systems we are primarily focussing on the locations of SR \citep[see][]{Mur1999, Lev2003, Bea2012a}, because SR are usually connected to highly eccentric motion, and so they can have a strong influence on a planet's habitability.

% -------------------------------------

% --- motivation ---
In this work we investigate a sample of binary star systems with a known giant planet. Our goal is to check how frequently do SR affect the orbits of bodies in the HZ and what are the conditions for this to happen. We employ a semi-analytical method \citep{Pil2016}, which allows to quickly identify SR and to exclude unstable regions from a stability analysis. A possible application of the method is for newly discovered exoplanetary systems of binary stars, where it would help to exclude those systems that cannot host additional planets in the HZ. This, in turn, makes the method especially interesting as a guide to select such binary star systems for in-depth observations that would allow planets in the HZ from a dynamical point of view.

% -------------------------------------

% --- article structure ---
Section \ref{sec:systems} presents the binary star systems selected for our investigation, and explains how we determined the habitable zones for those systems. In section \ref{sec:methods} we give a short summary of the semi-analytical method introduced by \citet{Pil2016} that allows to calculate accurately the location of SR in binary star systems. We present the results of the survey in section \ref{sec:results}, and we discuss the semi-analytical method and compare its performance relative to some purely analytical models in section \ref{sec:discussion}. Finally, in section \ref{sec:conclusions}, we summarise the results and give the conclusions.

% -----------------------------------------------------------------------------

% -----------------------------------------------------------------------------

% section "The systems"
% -----------------------------------------------------------------------------
% FILE    : systems.tex
% PURPOSE : binary star article, section "The systems"
% DATE    : 22 Nov 2016
% -----------------------------------------------------------------------------

\section{The Investigated Systems}\label{sec:systems}

% -----------------------------------------------------------------------------

% --- selection of systems ---
\edit{As of October 2016 in total 79 binary star systems with 113 planets had been discovered, of which 58 are of S-type and 21 of P-type}\footnote{\url{http://www.univie.ac.at/adg/schwarz/multiple.html}}.

For this survey we have selected S-type binary stars with (projected) separations $a_{B} \leq 100$~au; such binaries have typical orbital periods below 1000 years. This limit is based on the observation that planets in `wide' binaries (with $a_{B} \geq 300$~au) have essentially the same mass and period distributions as planets orbiting single stars \citep[see][]{Des2007}. Another consideration was that the distance ratio planet/secondary star should not become too small, as we are focusing on secular effects and a small ratio would lead to only weak effects (see section \ref{sec:methods}).

Given the above constraints on stellar separation, our sample consisted of 15 candidates. We discarded the Kepler-296 \citep{Rowe2014, Bar2015} and Kepler-444 \citep{Cam2015} systems, because they are compact multiplanetary systems with 5 planets. We also rejected OGLE-2008-BLG-092L \citep{Pol2014} and OGLE-2013-BLG-0341L \citep{Gou2014} due to insufficient data regarding the orbits of the companion star and the planet.

% -----------------------------------------------------------------------------

\subsection{System properties}

\begin{table*}
  \caption{List of investigated binary star systems ordered by decreasing stellar separation $a_{B}$. The planet hosting star is indicated by the capital letter $A$ or $B$ following the system's designation. Stellar masses $M_{A}$ and $M_{B}$ are in units of the solar mass, while planet masses $M_{P}$ are given in Jupiter masses.}\label{tab:binsys}
  \begin{tabular}{ l *{8}{r} }
    \hline

%                  & \multicolumn{3}{c}{Masses}              & \multicolumn{2}{c}{Semi-major axes} & \multicolumn{2}{c}{Eccentricities} & \\
  Name            & $M_{A}$      & $M_{B}$      & $M_{P}$   & $a_{B}$         & $a_{P}$           & $e_{B}$ & $e_{P}$                  & Reference \\
                  & $(M_{\sun})$ & $(M_{\sun})$ & $(M_{J})$ & (au)            & (au)              &         &                          & \\

    \hline

  94~Cet~Ab       & 1.34               & 0.90            & 1.68\phantom{0}  % masses
                  & 100\phantom{.0}    & 1.42                               % semi-major axes
                  & 0.26\phantom{$^a$} & 0.3\phantom{00}                    % eccentricities
                  & \citet{May2004,Roe2012} \\                              % references

  HD~177830~Ab    & 1.37               & 0.23            & 1.49\phantom{0}  % masses
                  & 97\phantom{.0}     & 1.22                               % semi-major axes
                  & 0.20$^a$           & 0.001                              % eccentricities
                  & \citet{Vogt2000,Rob2015} \\                             % references

  HD~177830~Ac    & \ldots{}           & \ldots{}        & 0.15\phantom{0}  % masses
                  & \ldots{}           & 0.51                               % semi-major axes
                  & \ldots{}           & 0.349                              % eccentricities
                  & \\                                                      % references

  GJ~3021~Ab      & 0.90               & 0.13            & 3.37\phantom{0}  % masses
                  & 68\phantom{.0}     & 0.49                               % semi-major axes
                  & 0.20$^a$           & 0.51\phantom{0}                    % eccentricities
                  & \citet{Naef2001,Cha2007} \\                             % references

  $\tau$~Boo~Ab   & 1.30               & 0.40            & 4.13\phantom{0}  % masses
                  & 45\phantom{.0}     & 0.05                               % semi-major axes
                  & 0.20$^a$           & 0.02\phantom{0}                    % eccentricities
                  & \citet{But1997,Rod2012} \\                              % references

  HD~126614~Ab    & 1.15               & 0.32            & 0.38\phantom{0}  % masses
                  & 36.2               & 2.35                               % semi-major axes
                  & 0.50$^b$           & 0.41\phantom{0}                    % eccentricities
                  & \citet{How2010} \\                                      % references

  $\alpha$~Cen~Bb & 1.11               & 0.93            & 0.004            % masses
                  & 23.4               & 0.04                               % semi-major axes
                  & 0.52\phantom{$^a$} & 0.0\phantom{00}                    % eccentricities
                  & \citet{Dum2012,Endl2015} \\                             % references

  HD~41004~Ab     & 0.70               & 0.40            & 2.54\phantom{0}  % masses
                  & 23\phantom{.0}     & 1.64                               % semi-major axes
                  & 0.20$^a$           & 0.2\phantom{00}                    % eccentricities
                  & \cite{Zuc2004} \\                                       % references

  HD~196885~Ab    & 1.33               & 0.45            & 2.98\phantom{0}  % masses
                  & 21\phantom{.0}     & 2.60                               % semi-major axes
                  & 0.42\phantom{$^a$} & 0.48\phantom{0}                    % eccentricities
                  & \citet{Cor2008,Cha2011} \\                              % references

  $\gamma$~Cep~Ab & 1.40               & 0.41            & 1.85\phantom{0}  % masses
                  & 20.2               & 2.05                               % semi-major axes
                  & 0.41\phantom{$^a$} & 0.05\phantom{0}                    % eccentricities
                  & \citet{Hat2003,Endl2011} \\                             % references

  Gliese~86~Ab    & 0.83               & 0.49            & 4.01\phantom{0}  % masses
                  & 19\phantom{.0}     & 0.11                               % semi-major axes
                  & 0.40\phantom{$^a$} & 0.05\phantom{0}                    % eccentricities
                  & \citet{Que2000,Fuh2014} \\                              % references

  Kepler-420~Ab   & 0.99               & 0.70            & 1.45\phantom{0}  % masses
                  & 5.3                & 0.38                               % semi-major axes
                  & 0.31\phantom{$^a$} & 0.772                              % eccentricities
                  & \citet{San2014} \\                                      % references    

    \hline

  \multicolumn{9}{l}{Notes: $^a$ assumed eccentricity; $^b$ upper limit for stable planetary motion} \\
  \end{tabular}
\end{table*}

\begin{figure}
  \includegraphics[width=\columnwidth]{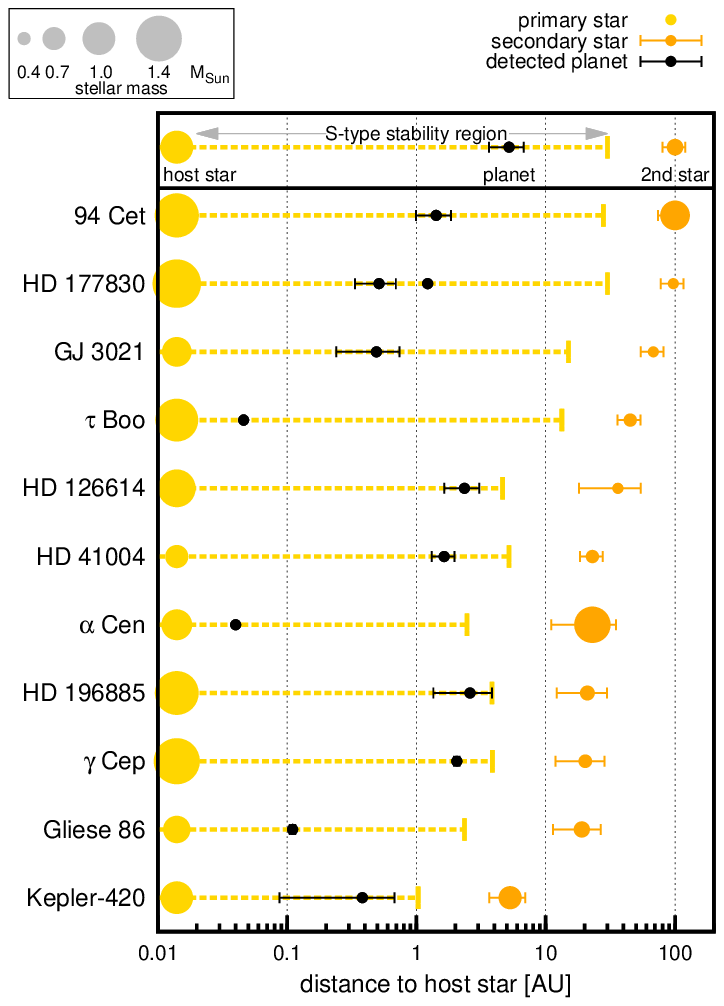}
  \caption{Overview of the binary star systems. The sizes of the circles representing the primary and secondary star indicate their masses according to the scale in the top left corner; the planet is not to scale. Error bars show the minimum and maximum distances of the respective objects, which are related to the eccentricity. The large dashed line marks the region of orbital stability for S-type motion about the host star in the respective system.}
  \label{fig:overview}
\end{figure}

% --- describe table 1 ---
In the end, 11 systems remained to be investigated. \autoref{tab:binsys} provides an overview of their basic parameters such as stellar masses ($M_{A}$, $M_{B}$) and the planet's minimum mass ($M_{P} = m \sin i$), as well as the semi-major axes ($a_{B}$, $a_{P}$) and eccentricities ($e_{B}$, $e_{P}$) for the secondary star and the planet, respectively. This table, in connection with \autoref{fig:overview}, summarises the main physical and dynamical aspects of the selected systems. The table includes references to the original detection paper, and in several cases to additional sources that provide important updates to the parameters (e.g. on stellar or planet masses, distances, etc.). For several binaries the eccentricity of the secondary star is unknown, in those cases we assume a moderate value of $e_{B} = 0.2$ based on \citet{Duq1991} and \citet{Rag2010}.

% --- describe figure 1 ---
\autoref{fig:overview} shows a graphical overview of the binary star systems. The circles for the primary (host) and secondary star indicate their masses (cf. \autoref{tab:binsys}); the box in the top left corner displays four typical cases for M-, K-, G-, and F-type stars. A large dashed line extending from the left marks the host star's region of orbital stability for planetary motion, beyond which it can no longer host S-type planets \citep{Hol1999, Pil2002}. As opposed to this, the secondary star might harbour S-type planets in a subregion outside this zone (not shown). We indicate the eccentricities of the secondary star and the planet with error bars having the colour of the respective object.
%% Finally, the location and extension of the habitable zone (HZ) is displayed by the coloured boxes.

% --- additional notes ---
Note that for $\alpha$ Centauri the existence of the planet is disputed. \citet{Dum2012} first claimed the discovery of a close-in planet with a few Earth masses. However, \citet{Hat2013} cautioned that the radial velocity signal may be induced by stellar activity rather than a planet. Recently \citet{Raj2016} presented evidence for a false positive detection. We also caution that in the Kepler-420 (KOI-1257) system the giant planet and secondary star have such high eccentricities \citep{San2014} that the dynamically stable zone around the host star is most probably devoid of additional planets.

% -----------------------------------------------------------------------------

\subsection{Habitable zone}

The classical habitable zone was defined for a single Earth-like planet on an almost circular orbit around a single star \citep{Kas1993, Kop2013}. Various attempts have been made to extend the concept of the HZ to binary stars. The major difference is to take into account the insolation onto the planet from both stars as a function of their luminosity, orbital parameters, as well as the changes in the planet's orbit.

%\subsubsection{Binary star HZ}
Here we follow \citet{Eggl2012} to calculate the so called Averaged Habitable Zones (AHZ), where the effective insolation onto the planet is on average within the prescribed maximum and minimum habitable insolation limits: $S_{\mathrm{in}} \geq \langle S_{\mathrm{eff}} \rangle_{t} \geq S_{\mathrm{out}}$. The AHZ represents an optimistic estimate of the HZ; besides the AHZ the Extended HZ (EHZ) and Permanent HZ (PHZ) pose more restrictive conditions on the effective insolation. In the following we always mean AHZ when using the term HZ. Alternative methods for calculating habitable zones in binary star systems were proposed by \citet{Kal2013} and \citet{Cun2014}. \edit{In contrast to \citet{Eggl2012}, these methods neglect variations of the planet's orbit and would yield slightly different results.}

\begin{table*}
  \caption{Extension of the binary star habitable zone (AHZ) and the dynamically stable zone (DSZ) with inner and outer borders, outer limit of the stability region, and location of secular resonances in the binary star systems. Dots indicate non-existent values.}\label{tab:hzlim}
  \begin{tabular}{ l *{6}{r} }

    \hline

  System       & \multicolumn{2}{c}{AHZ (au)} & \multicolumn{2}{c}{DSZ (au)} & Stability limit & Resonance location \\
               & inner & outer                & inner & outer                & (au)            & (au) \\

    \hline

  94~Cet       & 1.84  & 3.11  & 2.62   & 3.11   & 27.90 & \ldots     \\
  HD~177830    & 1.63  & 3.72  & 2.20   & 3.72   & 30.02 & 2.02, 3.78 \\ % NOTE post-MS HZ
  GJ~3021      & 0.81  & 1.40  & 1.24   & 1.40   & 14.96 & \ldots     \\
  $\tau$~Boo   & 1.67  & 2.84  & 1.67   & 2.84   & 13.30 & 0.45       \\
  HD~126614    & 1.09  & 1.90  & \ldots & \ldots &  4.63 & 0.91       \\
  $\alpha$~Cen & 0.72  & 1.26  & 0.72   & 1.26   &  2.45 & 1.16       \\
  HD~41004     & 0.79  & 1.40  & 0.79   & 0.94   &  5.20 & 0.37       \\
  HD~196885    & 2.06  & 3.53  & \ldots & \ldots &  3.84 & 1.11       \\
  $\gamma$~Cep & 2.51  & 5.90  & \ldots & \ldots &  3.86 & 0.78       \\ % NOTE post-MS HZ
  Gliese~86    & 0.63  & 1.10  & 0.63   & 1.10   &  2.34 & \ldots     \\
  Kepler-420   & 0.95  & 1.67  & \ldots & \ldots &  1.02 & \ldots     \\

    \hline
%\multicolumn{9}{l}{Notes: All numbers in this table are in astronomical units (au).}

  \end{tabular}
\end{table*}

% --- describe table 2 ---
\autoref{tab:hzlim} summarises the extent of the binary star HZ for each system (in au) by specifying the inner and outer border of the AHZ. Besides the HZ, we define the dynamically stable zone (DSZ) to identify those parts of the HZ where terrestrial planets \edit{can remain on long-term stable orbits subject to the gravitational perturbations of the giant planet and secondary star.} This table also indicates the border of the region of orbital stability for S-type motion. We have determined the boundary for regular S-type motion around the host star for each individual system by using the Fast Lyapunov Indicator \citep[FLI,][]{Fro1997} as presented in \citet{Pil2002}. These values are somewhat more restrictive than what the formula of \citet{Hol1999} would predict, as they also account for the planet's eccentricity. The last column shows the location of the linear secular resonance with the detected planet (see section \ref{sec:results}); missing entries for 4 systems mean that there is no such resonance.

\begin{figure}
  \includegraphics[width=\columnwidth]{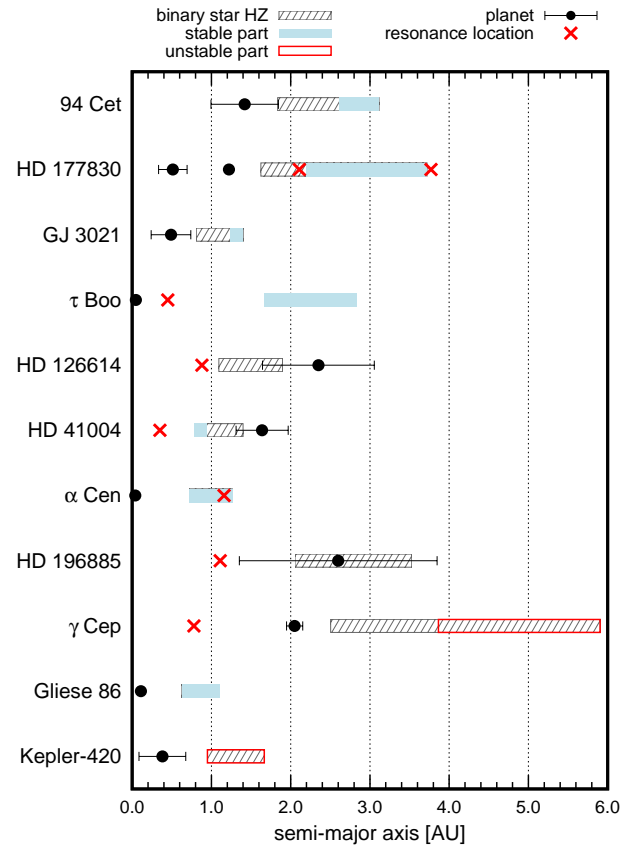}
  \caption{Binary star HZ for the planet hosting star (hatched boxes). The blue area is the intersection of the HZ with the dynamically stable region (see text for details). The two red-framed regions are beyond the stability limit. Black dots with horizontal error bars show the position and eccentricity of the planet, a red cross indicates the location of a (linear) SR.}
  \label{fig:habzone}
\end{figure}

% --- describe figure 2 ---
\edit{
In \autoref{fig:habzone} we plot the binary star HZ as hatched boxes. The intersection of the HZ with the region of stable orbital motion (DSZ) is plotted as the solid filled area, which we derived from numerical computations with test particles. The black dots show the location of the detected planets, where the error bars (extending from periastron to apoastron) visualise the planet's eccentricity. Interactions of a planet in the HZ with the giant planet can lead to dynamical instability in parts of the HZ by effects such as mean-motion and secular resonances. In several cases low-order MMRs with the giant planet occur at various locations inside the HZ; \edit{these MMR are not included in \autoref{fig:habzone} but collected in \autoref{tab:mmrloc}}. A cross symbol marks the location of linear secular resonances (see section \ref{sec:results}).}
For the systems $\gamma$ Cep and Kepler-420 parts of the HZ lie outside the S-type stable region (cf. \autoref{tab:hzlim}), this is indicated by a red border around the corresponding part of the HZ. For HD~177830 the two planets cause secular resonances at least at two locations, these are marked in \autoref{fig:habzone} and given in \autoref{tab:hzlim}.

\begin{table}
  \caption{Locations of mean-motion resonances inside the HZ of the investigated systems or nearby the SR. We indicate MMR by the test planet to giant planet period ratio.}\label{tab:mmrloc}
  \begin{tabular}{ l r@{:}l r }

    \hline

  System       & \multicolumn{2}{c}{MMR} & Resonance location (au) \\ \hline

  94~Cet       & 2&3  & 1.86 \\
               & 1&2  & 2.25 \\
               & 2&5  & 2.61 \\ \hline

  HD~177830~b  & 2&3  & 1.60 \\
               & 1&2  & 1.94 \\
               & 8&17 & 2.02 \\        % SR
               & 2&11 & 3.81 \\        % SR
  HD~177830~c  & 1&8  & 2.05 \\ \hline % SR

  GJ~3021      & 1&2  & 0.78 \\
               & 1&3  & 1.02 \\
               & 1&4  & 1.23 \\ \hline

  HD~126614    & 4&1  & 0.93 \\        % SR
               & 3&1  & 1.13 \\
               & 2&1  & 1.48 \\ \hline

  HD~41004     & 9&1  & 0.38 \\        % SR
               & 2&1  & 1.03 \\
               & 3&2  & 1.25 \\ \hline

  $\gamma$~Cep & 4&1  & 0.81 \\        % SR
               & 3&4  & 2.48 \\
               & 2&3  & 2.69 \\
               & 1&2  & 3.25 \\

    \hline

  \end{tabular}
\end{table}

% --- describe table 3 ---
\edit{\autoref{tab:mmrloc} provides a non-exhaustive list of mean-motion resonances with the giant planet(s) that happen to fall inside the HZ or nearby the SR. The numbers in the second column indicate the orbital period ratios; when the latter number is larger there is an exterior resonance ($a_P < a$), while it is smaller for interior resonances ($a < a_P$). We note that some of the MMR are located close to the border of the HZ or coincide with the transition from unstable to regular motion. For HD~177830, HD~126614, HD~41004, and $\gamma$~Cephei we included those MMR that are nearest to the location of the SR (cf. \autoref{fig:habzone} and \autoref{tab:hzlim}).}

% --- post main-sequence stars ---
%\subsubsection{Post-main-sequence stars}
We note that HD~177830~A and $\gamma$~Cephei~A are post-main-sequence stars \edit{(subgiants)}, and the companion of Gliese~86 is a white dwarf \citep{Fuh2014}. In these cases we cannot calculate the HZ as for ordinary main sequence stars. For Gliese~86 a resort is to ignore the contribution of the white dwarf (WD) companion, and to approximate the current HZ by the host star's HZ alone. This is justifiable by the low luminosities of cool WDs, e.g. $L/L_{\sun} < 10^{-3}$ for $T_{\mathrm{eff}} < 10^{4}$~K \citep[see][]{Lie2005}. For the other two evolved stars we use the model of \citet{Ram2016} to calculate the HZ of post-main-sequence stars. Their model provides the effective stellar flux $S_{\mathrm{eff}}$ for the inner and outer boundary of the HZ for a grid of stellar masses between $0.5-1.9 \, M_{\sun}$, following the stellar evolution from the beginning of the Red Giant Branch until the end of the Asymptotic Giant Branch.
%\edit{Note that this model does not take into account planetary orbit variations (e.g. due to stellar mass loss).}
We determine the effective stellar flux values for HD~177830 and $\gamma$~Cep by systematic interpolation between the masses of stars in the model to fit the current masses, which yields the values in \autoref{tab:hzlim} when combining the results with the published luminosities \citep{Jof2015}.

% --- main-sequence HZ models ---
In addition, we tried to recover the main-sequence parameters for the two stars in order to obtain an estimate of their main-sequence HZ. A main-sequence mass for $\gamma$~Cep and HD~177830 can be calculated through a Bayesian estimation of stellar parameters, i.e. calculating the most likely values of stellar mass, age, and radius, knowing the effective temperature, the metallicity, the $V$-photometry, and the distance (parallax), see \citet{Sil2006} for details. The assumed Bayesian priors for the calculations are an Initial Mass Function for single stars from \citet{Cha2001} (log-normal), and a constant Star Formation Rate in the age interval from $0.1 - 12.0$~Gyr. For the Bayesian analysis we adopted the values from \citet{Sou2016} for $T_{\mathrm{eff}}$, [Fe/H], and $V$-magnitude, and we used the parallax from \citet{Van2007}. Given the wide range of possible metallicities and effective temperatures for $\gamma$~Cep, as well as the varying mass estimates \citep[see][]{Tor2007}, we can only set an upper limit on the main-sequence mass of $M = 1.62 \pm 0.10$~$M_{\sun}$. For HD~177830 we find $M = 1.41 \pm 0.03$~$M_{\sun}$, which is consistent with the mass given by \citet{Rob2015} to within the error bars. We use the \citet{Kop2013} model to calculate HZ boundaries for runaway/maximum greenhouse, where we assume both stars to be of spectral type F0 and having an effective temperature of 7200~K (the upper limit in the model). This results in HZ distances of $1.86 - 3.20$~au for $\gamma$~Cep, and $1.77 - 3.05$~au for HD~177830. However, given the many assumptions about metallicities, masses, and luminosities these ranges should be used with caution.

% -----------------------------------------------------------------------------

% -----------------------------------------------------------------------------

% section "The methods"
% -----------------------------------------------------------------------------
% FILE    : method.tex
% PURPOSE : binary star article, section "The methods"
% DATE    : 22 Nov 2016
% -----------------------------------------------------------------------------

\section{Methods}\label{sec:methods}

To investigate the secular dynamics of the binary star systems we use a semi-analytical method, that is presented and discussed in more detail in \citet{Pil2016}; here we give just a quick summary: We treat the binary star -- giant planet -- test planet four-body problem as two coupled three-body problems. First, for the pair giant planet -- secondary star, we derive from a single $N$-body numerical integration the secular precession frequency of the giant planet's pericenter with a characteristic frequency determination. This precession frequency is approximated as being fixed. Next, for the test planet -- giant planet pair, we can apply the Laplace-Lagrange perturbation theory to find the locations of secular resonances.

% -----------------------------------------------------------------------------

\subsection{Laplace-Lagrange theory}

The Laplace-Lagrange secular perturbation theory (LL) was developed to serve as a first approximation to the dynamical behaviour of the solar system \citep[see][]{Mur1999}. It is a first order theory in the masses, and includes terms in the eccentricity and inclination of degree $e^2$ and $(\sin i)^2$, while higher order terms are neglected. Nevertheless, this approximation already allows to determine the location of the most important linear secular resonances.

Let us consider an exoplanetary system consisting of a test planet (with mass $m \ll m_{j}$) moving under the influence of the host star ($m_{0}$), a giant planet ($m_{1}$), and the secondary star ($m_{2}$). All bodies are assumed to be point mass objects and located in a common plane.

To set up the equations of motion usually the variables $(h = e \sin \varpi, k = e \cos \varpi)$ are defined. The general solution to the equations of motion is then of the form
\begin{equation} \label{eq:tpsol}
  \begin{split}
    h(t) &= e_{\mathrm{free}} \sin(g t + \phi) - \sum_{j=1}^{2} \frac{\nu_{j}}{g - g_{j}} \sin(g_{j} t + \phi_{j}),\\
    k(t) &= e_{\mathrm{free}} \cos(g t + \phi) - \sum_{j=1}^{2} \frac{\nu_{j}}{g - g_{j}} \cos(g_{j} t + \phi_{j}).
  \end{split}
\end{equation}
Here $a, e, n, \varpi$ are the semi-major axis, eccentricity, mean motion, and longitude of pericenter of the test particle, respectively, and variables with subscript $j$ are associated to the perturbers ($j = 1$ the giant planet, $j = 2$ the secondary star). The constants $e_{\mathrm{free}}$ and $\phi$ are given by the initial conditions for the particle, while $g_{j}$, $\phi_{j}$, and $\nu_{j}$ follow from the secular dynamics of the massive perturbers.

The quantity $g$ is the test particle's proper secular frequency calculated from
\begin{equation} \label{eq:propfreq}
  g = \frac{n}{4} \sum_{j=1}^{2} \frac{m_{j}}{m_{0}} \alpha_{j} \bar{\alpha}_{j} b_{3/2}^{(1)} (\alpha_{j}).
\end{equation}
This frequency $g$ depends on the semi-major axes of the particle and of the perturbers via the ratio $\alpha_{j} = a / a_{j}$ (in case $a < a_{j}$) or $\alpha_{j} = a_{j} / a$ (when $a > a_{j}$), the Laplace coefficient $b_{3/2}^{(1)}(\alpha_{j})$, as well as on the perturber's mass $m_{j}$. Equations \eq{eq:tpsol} and \eq{eq:propfreq} can easily be extended to an arbitrary number $N$ of perturbing bodies. From equation \eq{eq:propfreq} we can see that $g \propto O ( \alpha^{3} )$ (using $b_{n}^{(k)} \propto O ( \alpha^{k} )$), so the secular frequency quickly diminishes with decreasing $\alpha$, i.e. when the perturber is located farther away. For this reason the main contribution to $g$ stems from the giant planet which is much closer than the secondary star, despite the difference in the masses. Extending this argument to well separated binaries (with $a_{B} > 100$~au) it becomes clear that a distant secondary star would have only a negligible influence on the test planet's secular frequency. In that case the system can be approximated as a single star with a giant planet having a fixed forced secular frequency $g_{1}$.

Wherever $g \approx g_{j}$ in equation \eq{eq:tpsol}, the particle's secular frequency equals one of the perturber's frequencies, which then gives rise to a secular resonance. In these cases the particle suffers from large secular perturbations due to the perturber $j$, where the particle's forced eccentricity can reach values close to $e = 1$. Then, it can cross the orbit of the giant planet or be ejected from the system by a close approach to the host star.

% -----------------------------------------------------------------------------

\subsection{Numerical part}

In equation \eq{eq:tpsol} we need to know the frequencies $g_{j}$ accurately in order to determine the location where $g - g_{j} \approx 0$, i.e. the location of a linear secular resonance. We chose to determine the fundamental frequencies numerically throughout this study. This is necessary as the eccentricities of the binary stars are large for all selected system ($e_B \ge 0.2$, see \autoref{tab:binsys}), such that a low-order analytical theory (like the LL theory) has only a limited accuracy.

We use the Lie-series method \citep{Han1984, Eggl2010, Ban2012} and the Mercury integrator's Radau method \citep{Cha1999, Eve1974} as complementary tools to integrate the equations of motions of the selected binary star-planet system. Next, we perform a frequency analysis on the results of our integrations to extract the fundamental frequencies; this step is accomplished by using the Fast-Fourier Transform (FFT) library FFTW\footnote{\url{http://fftw.org/}} of \citet{Fri2005}. As an independent check we use the tool SigSpec of \citet{Ree2007} that performs a Discrete Fourier Transform (DFT) and gives the probability that a peak is not due to noise in the time-series. Whenever possible we remove short period contributions (associated with the orbital frequencies of the giant planet and the secondary star) by a low-pass filter. From the frequency analysis we obtain the giant planet's secular frequency $g_{P}$ (or $g_{1}$), which is effectively the time-averaged frequency $\langle g(t) \rangle_{T}$ over the integration time $T$. Finally, to determine the locations of secular resonances for the test particles, we identify the intersection points of equation \eq{eq:propfreq} with the values $g_{j}$.

\begin{figure}
  \includegraphics[width=\columnwidth]{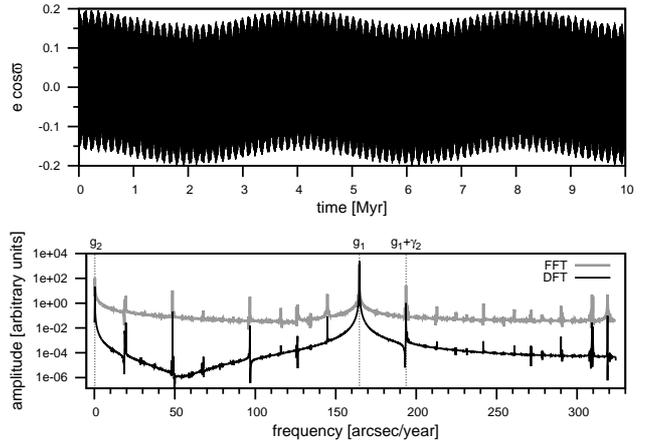}
  \caption{Comparison of DFT and FFT results. Top: time evolution of variable $k(t)$ (from eq. \eqref{eq:tpsol}) for the system HD~41004. Bottom: Fourier amplitudes for the signal from above. The peaks identified as $g_1$, $g_2$ denote the secular frequencies of the planet and the star, respectively.}
  \label{fig:dftfft}
\end{figure}

% --- describe figure 3 ---
\edit{\autoref{fig:dftfft} shows an example for the processing of the numerical integrations. The upper panel displays the time evolution of $k(t)$ during the first $10^7$~years from a 30~Myr integration of the system HD~41004. In the bottom part we see the DFT (black) and FFT (grey) spectrum of the full signal, where frequencies above 330~arcsec~yr$^{-1}$ have been filtered out. The two curves are shifted vertically relative to each other, which means that their amplitudes were scaled differently and are not directly comparable; however the horizontal frequency axis is the same for both. Both techniques agree excellently in the identification of the peaks, where $g_1$ and $g_2$ are the system's main secular frequencies attributable to the giant planet and the secondary star. The third largest peak $g_{1} + \gamma_{2}$ is a linear combination of the planet's secular frequency with a short period term from the star's signal, but we have not attempted to identify its origin. In the top panel the effect of $g_2$ is clearly visible as the $\sim 4$~Myr oscillation, while $g_1$ is superposed by a longer period.}

% -----------------------------------------------------------------------------

\subsection{Relativistic effects}

From \autoref{tab:binsys} we can observe that there are several planets with $a_{P} \lesssim 0.5$~AU. For such close-in planets the general relativistic precession of the perihelion (GRP) becomes important, like in the case of Mercury in our solar system \citep{Las2008}. We determine this additional precession frequency for the longitude of pericenter in the simplified form taken from \citet{Bra2009}:
\begin{equation} \label{eq:relprec}
  \left\langle \frac{\mathrm{d} \varpi}{\mathrm{d} t} \right\rangle = \frac{3 G M}{c^2} \frac{n_P}{a_{P} (1 - e_{P}^{2})}.
\end{equation}
As a check of the formula we performed direct numerical test calculations of the relativistic equations of motion, solving them either in the full Einstein-Infeld-Hoffmann form \citep{Beu2005, Bru2007}, or in a more approximate and simplified variant \citep{Ban2012}. Results from these calculations agree well with equation \eq{eq:relprec}, so that we use it instead of a more time-consuming numerical integration. We do not include tidal effects between the close-in planets and the host star \citep[see][]{Fab2007}, because the orbits are already quite circular, such that a further circularisation is not expected.

\begin{figure}
  \includegraphics[width=\columnwidth]{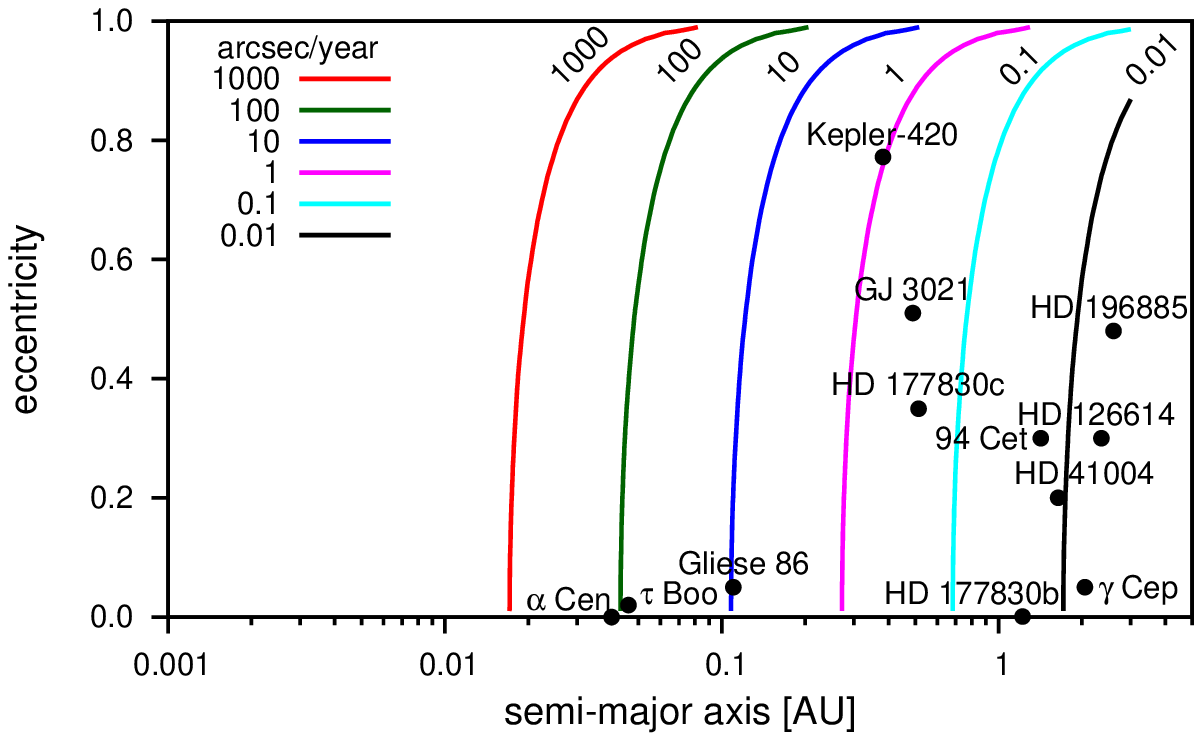}
  \includegraphics[width=\columnwidth]{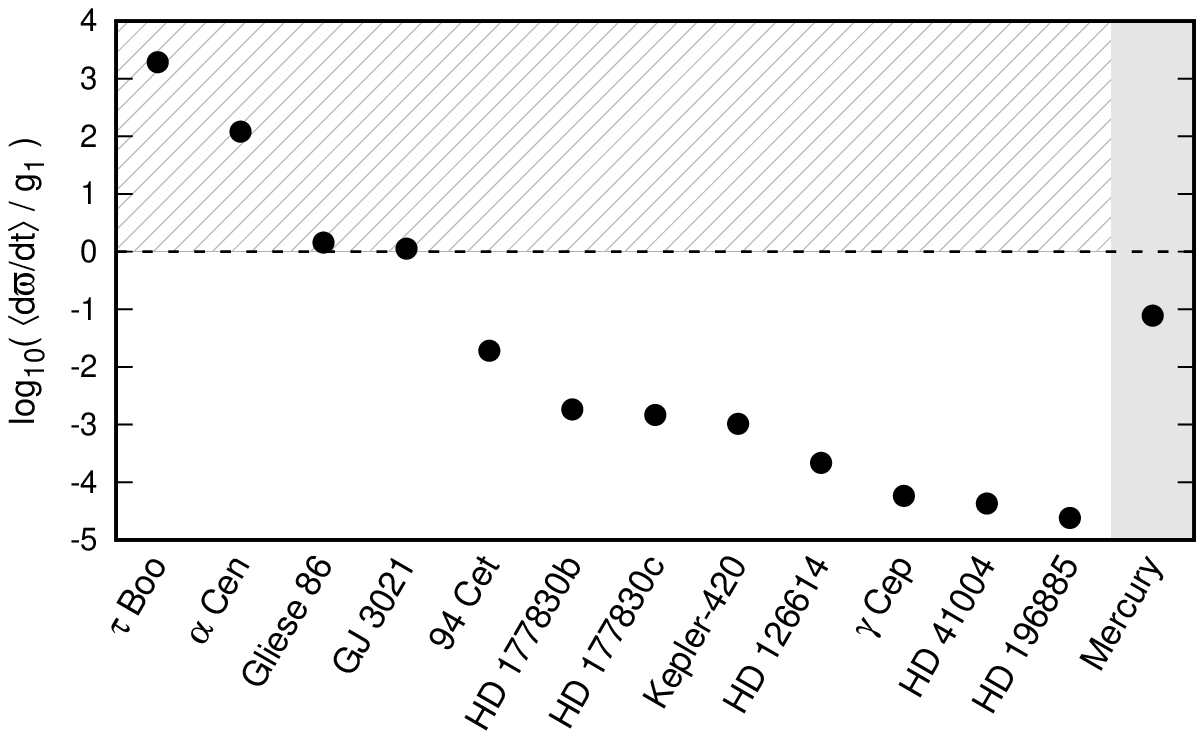}
  \caption{Top: Contour levels indicating the order of magnitude of the giant planet's general relativistic precession of perihelion (in arcseconds per year) for different binary star systems. Bottom: Ratio of the relativistic precession of perihelion relative to the secular precession frequency induced by the binary companion. In the shaded area the relativistic precession frequency is at least as important as the secular one.}
  \label{fig:relprec}
\end{figure}

% --- describe figure 3 ---
\autoref{fig:relprec} shows the absolute and relative magnitude of the GRP. The upper panel plots contour levels for equation \eq{eq:relprec} in the $a_{P} - e_{P}$ parameter space. The exoplanets are placed according to their parameters in \autoref{tab:binsys}. In the lower panel of this figure we plot the ratio of the GRP to the secular precession frequency; for comparison Mercury is also included. Whenever the GRP is of the same order of magnitude or larger than the secular frequency of the giant planet, which is indicated by the dashed line and shaded area, we include the general relativistic effect. This figure demonstrates that the GRP can be orders of magnitude larger than the precession frequency induced by purely Newtonian interaction between the giant planet and secondary star. In case of $\tau$~Boo, $\alpha$~Cen, Gliese~86, and GJ~3021---lying above the dashed line---we have to include this additional contribution.

% -----------------------------------------------------------------------------

% -----------------------------------------------------------------------------

% section "Results"
% -----------------------------------------------------------------------------
% FILE    : results.tex
% PURPOSE : binary star article, section "Results"
% DATE    : 22 Nov 2016
% -----------------------------------------------------------------------------

\section{Results}\label{sec:results}

It becomes evident from \autoref{tab:binsys} that HD~177830 is the only binary with two planets, in all other cases only a single planet has been detected so far. This leads us to the following questions: Can there be yet undetected additional (Earth-like) planets? Would their radial velocity (RV) signals be detectable with current detectors?
 
% -----------------------------------------------------------------------------

% --- existence of additional planets ---
\subsection{Observational aspects for finding additional planets}

The lack of known Earth-mass planets in the selected binary star systems is a direct consequence of their detection via the RV method (except for Kepler-420). Current RV planet search programs have typical measurement precisions between $0.8 - 5.0$~\mps{} \citep{Fis2016}, which favours the detection of massive and/or close-in planets. The discovered exoplanets from \autoref{tab:binsys} produce RV signals of $5 - 500$~\mps{} with rms fit residuals of $4 - 20$~\mps{} \citep[see][Table 3]{But2006}. In contrast, Earth-like and super-Earth-like planets with masses of 1~$M_{\earth}$ and 10~$M_{\earth}$ at a typical distance of 1~au would produce signals of 0.09 and 0.9~\mps{}, respectively.

We estimate the maximum RV signal of a terrestrial planet using the approach of \citet{Eggl2013}. This method assumes a planet on an initially circular orbit. Perturbations from both the secondary star and the giant planet cause its eccentricity to vary secularly (i.e. over long periods of time) between $e = 0$ and $e = \emax$, where the RV signal will be maximal at the latter value:
\begin{equation} \label{eq:radvel}
  V_{r}^{\mathrm{max}} = M_{P} \sin i \sqrt{\frac{G}{a (M_{A} + M_{P})}} \sqrt{\frac{1 + \emax}{1 - \emax}}.
\end{equation}
\citet{Eggl2012} used an analytic formula due to \citet{Geo2003} to calculate the value of $\emax$ depending on the masses and orbital parameters of the perturbers. Combining these methods we calculate that planets with 1~$M_{\earth}$ in the HZ of the respective host star never exceed a RV semi-amplitude of 1~\mps{}, while planets of 10~$M_{\earth}$ can reach 2~\mps{} only in the innermost part of the HZ.

Consequently, additional terrestrial planets can be hidden in the RV signal residuals and could be detected in the future with more precise measurements. We are aware that very tight binaries can pose a problem as the gas giant might have interfered with the formation of terrestrial planets \citep{Qui2007}. Nevertheless, \citet{Ray2005, Fogg2007, Ogi2014} presented promising results that show the feasibility of terrestrial planet formation after migration of a giant planet to orbital distances $\leq 0.5$~au. Although these simulations covered only disks of single stars, the accretion of a few terrestrial mass objects outside the hot Jupiter's orbit should be viable even for truncated disks of binaries \citep{Hag2007}.

% --- Neptune / Jupiter mass planets ---
\edit{Another possibility is the existence of Neptune to Jupiter mass planets at larger orbital distances. Based on the stability limits (cf. \autoref{tab:hzlim} and \autoref{fig:overview}) such planets could be imagined for the wider binaries, but excluding HD~126614, HD~196885, and Kepler-420. However, Jupiter mass planets up to a reasonable distance of about 5~au (for reasons of the long orbital periods) would always produce RV signals of more than 10~\mps{}, and hence this kind of planets should have already been detectable so far. Neptune mass planets (with about 20~$M_{\earth}$) are more difficult to detect and would produce signals of a few \mps{}. Anyway, the presence of an additional perturber increases the number of fundamental frequencies and leads to more possible combinations that can become resonant (also the periods become shorter, see discussion on HD~177830 in section \ref{sec:intext}). Therefore, the main focus of this survey is on a single giant planet in binary star systems; a similar study for two planets in binaries is work in progress.}

% -----------------------------------------------------------------------------

% --- massless vs massive planets ---
\subsection{Testing massless versus massive planets}

It is common practice to use the test particle approximation for the dynamics of low mass objects. This approximation is suitable for small mass-ratios, like for asteroids to Jupiter or a terrestrial mass planet to a star. Here we demonstrate that is is also applicable in the case of terrestrial planets in binary star systems interacting with a giant planet.

\begin{figure}
  \includegraphics[width=\columnwidth]{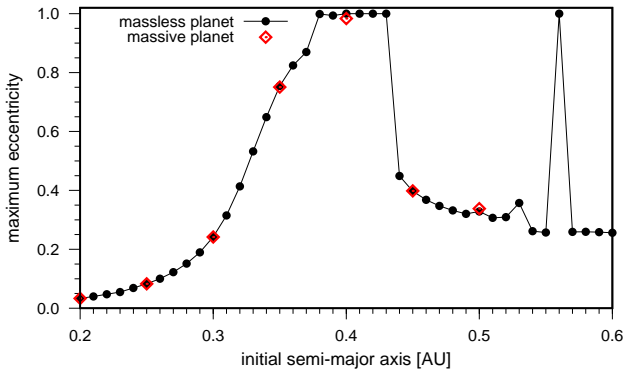}
  \caption{Maximum eccentricities of massless (black points) and massive (red diamonds) test planets in the system HD~41004, distributed in an interval around the location of the secular resonance.}
  \label{fig:maxecc}
\end{figure}

% --- describe figure 4 ---
\autoref{fig:maxecc} compares the restricted to the full four-body problem for HD~41004. The black points indicate the initial semi-major axes of massless test particles that start on initially circular orbits in an interval centered on the SR (at $a = 0.37$~au, see \autoref{tab:hzlim}). In the course of the numerical integration they reach a maximum eccentricity defined as
\begin{equation} \label{eq:maxecc}
  \emax = \max_{t \leq T} \left[ e(t) \right].
\end{equation}
The curve shows the effect of secular forcing of the eccentricity amplitude when approaching the SR. Particles located further away from the resonance remain on low to moderately eccentric orbits, while those in the vicinity of the resonance are ejected from the system ($\emax \ge 1$) in relatively short time. The spike at $a = 0.56$~au is due to the 5:1 MMR with the giant planet, and the neighbouring smaller spikes are caused by high-order MMR. The red diamonds represent the maximum eccentricity of Earth-mass planets. Each point was calculated in a separate run to avoid mutual interactions between these planets that could tamper with the results. It is visible that there is an excellent agreement between the massless and massive planet eccentricities.

% -----------------------------------------------------------------------------

\subsection{Interior versus exterior giant planets}\label{sec:intext}

% --- define internal & external GP ---
The planetary systems in \autoref{tab:binsys} can roughly be distinguished into two classes depending on the location of the giant planet. When the giant planet is exterior to the HZ, like Jupiter in the solar system, we refer to these systems as `exterior cases'. Otherwise, we refer to giant planets located interior to the HZ as `interior cases'. Regarding \autoref{fig:habzone} we have to clarify that we consider HD~196885 and $\gamma$~Cephei as being exterior cases, despite both planets being located exactly inside the (main-sequence) HZ.

\begin{figure*}
  \includegraphics[width=2.0\columnwidth]{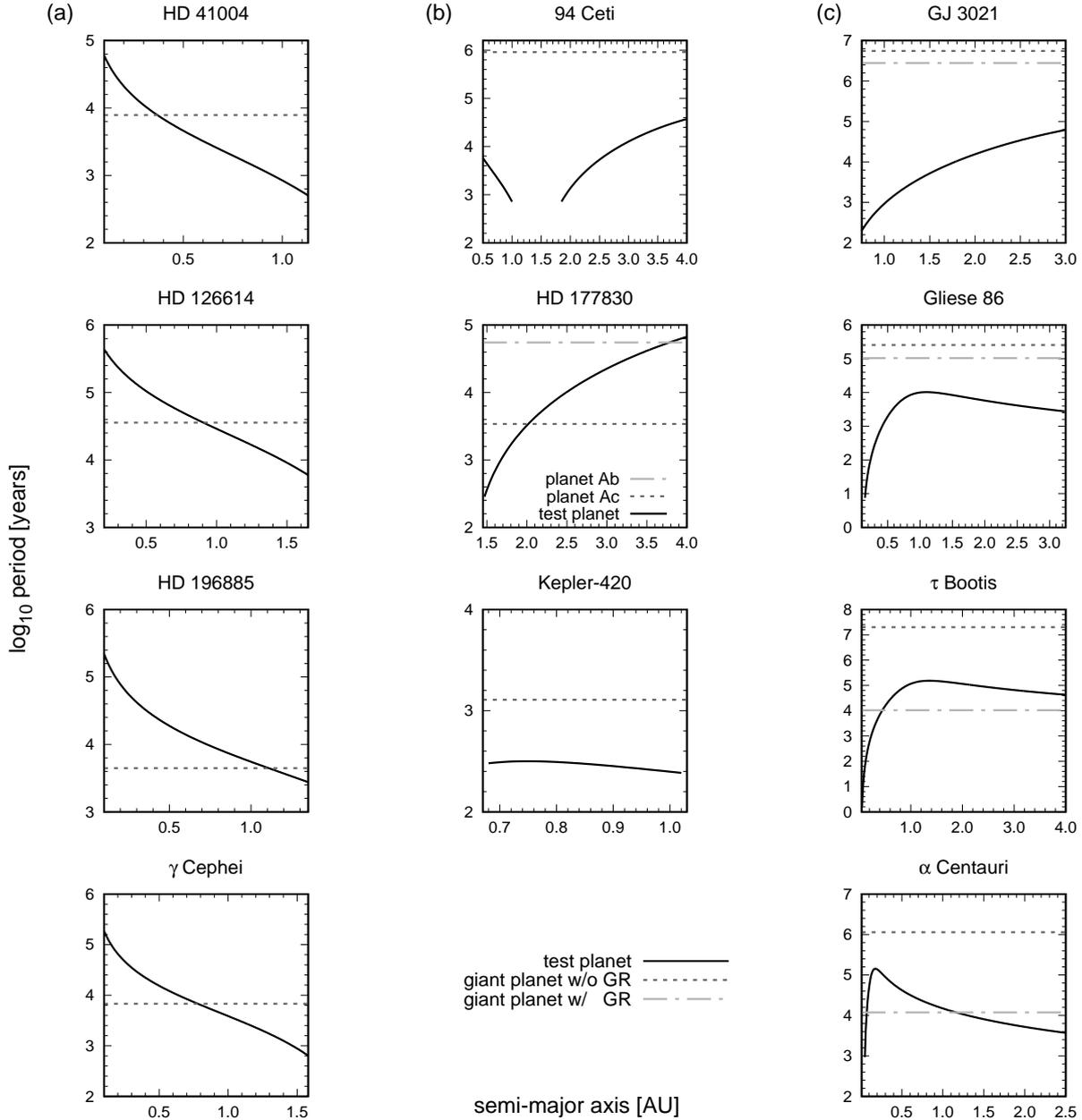}
  \caption{Plot of the proper secular periods of test planets as a function of their semi-major axis. An intersection of the black curve with the horizontal line indicates the occurrence of a linear secular resonance. The left column (a) collects systems with exterior giant planets, while columns (b) and (c) show systems with a giant planet that is interior to the orbit of the test planet. Column (c) presents cases where the close-in giant planet is affected by the general relativistic precession (dashed line without GR, dot-dashed line including GR).}
  \label{fig:periods}
\end{figure*}

% --- discuss figure 5 ---
\autoref{fig:periods} displays the proper secular period depending on the test planet's semi-major axis. The intervals are chosen in such a way, that the test particles never approach the giant planet closer than three Hill radii, or alternatively never closer than the pericenter and apocenter distance of the giant planet (for column (a) and for the other two columns, respectively), whichever is larger. The systems in column (a) all belong to the exterior planet type, and in each case there is a linear SR interior to the giant planet's orbit. Contrary to that, for the interior cases in columns (b) and (c) there is no linear SR, except for HD~177830. In the right column it is visible that the precession periods can decrease appreciably when including the relativistic precession of the perihelion, such that a SR with the close-in planet may occur.

% --- qualitative difference in dynamics ---
The results suggest that there is a qualitative difference in the dynamics depending on the location of the giant planet. In all of the exterior cases a single giant planet causes a linear SR at some place interior to its orbit. As opposed to this, a single interior giant planet does not cause a linear SR, at least for $a_{P} \ga 0.1$~au when the GRP is insufficient to increase the precession frequency by the required amount.

% --- WHY important + implications ---
This difference has important implications for the HZ and the possible presence of habitable planets. A hot/warm Jupiter in a binary star system would not cause a SR in the HZ. Planets in the HZ of Sun-like stars in such systems would only be subject to MMR, but the period ratio to the close-in planet only allows for high-order MMR that have a weak effect. The configuration of giant planets exterior to the orbit of terrestrial planets, based on the architecture of the solar system, might not be the optimal model for habitable planets. It is well known that Jupiter and Saturn are causing the $\nu_{5}$ and $\nu_{6}$ SR in the solar system, and the latter plays an important role for perturbing the main-belt of asteroids. In a similar way the accretion of planet-mass objects could be obstructed by exterior giant planets.

% --- SR and impacts + water delivery ---
\edit{Indeed, \citet{Ban2015, Ban2016} showed that SR can exist inside the main-belt analogues of certain binary star -- giant planet configurations. When the SR lies in the asteroid belt it can overlap with MMRs and be responsible for a high rate of asteroids crossing the HZ and, therefore, for an enhanced impact rate and water transport to the HZ (Bancelin, in preparation).
% TODO FIXME: 'they' --> \cite{bancelin16b, in prep}
In addition, they compared the water delivery to planets (or planetary embryos) orbiting in the HZ, according to the location of the SR---either inside the HZ or inside the asteroid belt. They concluded that:
\begin{enumerate}
  \item A planet moving on an eccentric orbit (when the SR lies inside the HZ) would increase the probability to collide with incoming wet asteroids. However, the relatively high impact velocity will strongly reduce the quantity of water finally arriving at the planet's surface.
  \item A planet moving on a nearly circular orbit (when the SR lies inside the asteroid belt) can boost its water content due to a higher flux of incoming wet asteroids at relatively low impact velocities.
\end{enumerate}
}

But what about multi-planet systems? Do these objections also apply to them? HD~177830 is an analogy to the solar system, just with the giant planets interior to the HZ. In that case there are two giant planets whose mutual gravitational interaction increases the apsidal precession frequencies relative to those that would be caused by the secondary star alone. For multiple giant planets SR are a more common phenomenon, and this leads to a large variety of possible planetary system architectures \citep[see][]{Lev2003}.

% --- physical interpretation + explanation ---
The explanation for the different behaviour of external and internal cases lies in the behaviour of the analytical curve described by equation \eq{eq:propfreq}. The secular frequency $g$ is the superposition of two functions of $\alpha$ that are strictly monotonic. In the exterior case the periods for the test planets are large nearby the host star for small values of $\alpha$, i.e. the torque is small and the exchange of angular momentum with the giant planet is inefficient. The periods decrease monotonically and approach zero as $a \rightarrow a_{P}$ (because of the properties of the Laplace coefficient $b_{3/2}^{(1)}(\alpha)$ when $\alpha \rightarrow 1$). At some point in between the curve must necessarily cross the value of $g_{1}$, which gives rise to the SR. For the interior case, on the one hand the periods are first low in the vicinity of the giant planet, but increase monotonically with increasing distance to the planet (i.e. $\alpha = a_{P} / a$ is decreasing). On the other hand, the periods are steadily decreasing under the influence of the secondary star. From the combination of these two components the secular periods rise to some peak value before they start decreasing again (see \autoref{fig:periods}). For the systems investigated the secular periods of the giant planet and the peak value for test planets are differing by more than a factor of ten. Consequently, interior giant planets are unable to cause a linear SR, except for when the GRP is large enough.

% -----------------------------------------------------------------------------

\subsection{Effect of secular resonances}

Now let us investigate two typical cases for the effect of the SR for exterior cases. For a more detailed analysis of the dynamical stability see e.g. \citet{Pil2010} and \citet{Funk2015}.

\begin{figure}
  \includegraphics[width=\columnwidth]{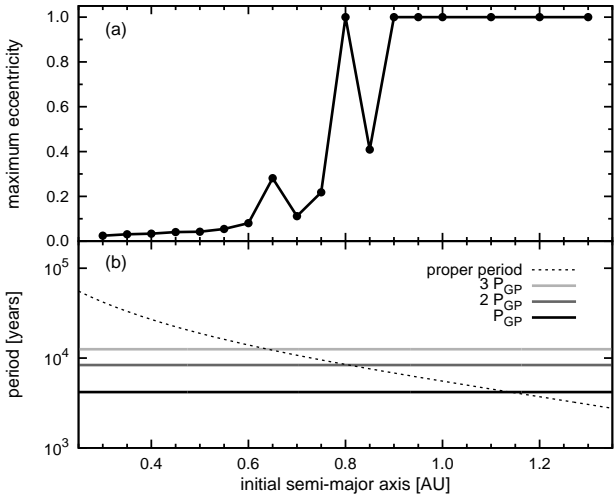}
  \caption{Effect of the secular resonance on test particles in the HD~196885 system. Over a time interval of $T = 10^{6}$ years the maximum eccentricity rises to unity (top) at the locations where the test particle's proper period (bottom; dashed black curve) is equal to multiples of the giant planet's secular period (horizontal lines).}
  \label{fig:hd196885}
\end{figure}

% --- describe figure 6 ---
\autoref{fig:hd196885} shows the maximum eccentricities of test particles reached over an integration time of $10^{6}$ years in panel (a). All test particles start on initially circular orbits, but those beyond $a \simeq 0.9$~au quickly reach a value of $e \ge 1$ and are ejected from the system by a close encounter with the giant planet. At two other locations we observe a local increase of the maximum eccentricity. These locations correspond to the non-linear\footnote{\edit{We follow \citet[][section 2]{Kne1991} for the definition of `linear' and `non-linear' secular resonances.}} secular resonances $2 g - g_{P} = 0$ and $3 g - g_{P} = 0$, while nearby there are also the 6:1 and 8:1 mean motion resonances at $a \approx 0.79$ and $a \approx 0.65$ AU, respectively. In the bottom panel (b) we show the intersection points of the test particle's proper secular period curve (dashed black) with multiples of the giant planet's period ($P_{GP}$; the horizontal lines). One can observe that the intersection points correspond to the peaks in maximum eccentricity. This demonstrates that the semi-analytical method is also useful and capable to detect non-linear SR. Note that for this simulation we have reduced the giant planet's initial eccentricity to $e_{P} = 0.3$, as otherwise the planet's perihelion distance ($q \approx 1.3$~au) and the resonance width of the 3:1 MMR (at $\sim 1.25$~au) would destabilise the majority of test particles in this region.

\begin{figure}
  \includegraphics[width=\columnwidth]{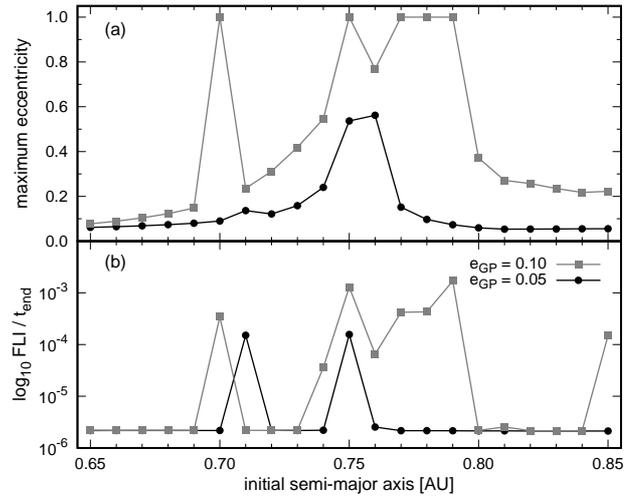}
  \caption{Effect of the secular resonance on test particles in the $\gamma$ Cephei system for two values of the giant planet's eccentricity (black and grey curves). Top: The maximum eccentricity as a function of the initial semi-major axis. Bottom: normalised FLI of the same region.}
  \label{fig:hd222404}
\end{figure}

% --- describe figure 7 ---
\autoref{fig:hd222404} shows the variation of maximum eccentricity (top) and Fast Lyapunov Indicator (FLI; bottom) as a function of the initial semi-major axis of test particles. In the first case the giant planet's eccentricity is $e_{P} = 0.05$ (black curve), which is actually close to the dynamically relaxed value of $e_{F} \approx 0.056$ for the forced eccentricity obtained by \citet{And2016} from an analytical model. Increasing the giant planet's eccentricity moderately to $e_{P} = 0.1$ has a strong effect on the test particles (grey curve). The maximum eccentricities now reach values sufficiently high for ejection from the system. The FLI is a chaos indicator that is sensitive to mean-motion resonances \citep{Fro1997}; a low FLI value means regular motion, while a high value indicates chaotic motion. Using the FLI the two peaks of the black curve in panel (b) are clearly identified as corresponding to the 5:1 ($a = 0.71$~au) and 9:2 ($a = 0.75$~au) MMR, respectively.

% --- overlap of MMR + SR --> Mudryk & Wu 2006, displacement of MMR subresonances ---
The main mechanism responsible for the strong increase of the maximum eccentricity is the forced eccentricity component, which depends on the perturber's eccentricity. A minor increase in the giant planet's eccentricity from 0.05 to 0.1 (cf. the black and grey curves in panel (a)) is sufficient to drive the forced eccentricity to values near unity. This means that already a moderately eccentric giant planet is able to perturb terrestrial planets quite effectively. The overlap of (relatively high-order) MMR and SR \citep{Wis1980} clearly leads to chaotic behaviour and effectively empties the region around the SR. Resonance overlap has also been studied by \citet{Mud2006} for binary systems, but in the case of the secondary star ejecting the exoplanet. They found that secular forcing of the companion shifts the centers of MMR, thus leading to extended resonance overlap regions. We can observe this effect in panel (b) for the 5:1 MMR that is shifted from 0.71~au (for $e_{P} = 0.05$) to 0.70~au (for $e_{P} = 0.1$), consistent with the calculations of \citeauthor{Mud2006}. This effect is less pronounced for the weaker 9:2 MMR, and would only be visible with a higher resolution in the grid of initial conditions.

% -----------------------------------------------------------------------------

% -----------------------------------------------------------------------------

% section "Discussion"
% -----------------------------------------------------------------------------
% FILE    : discuss.tex
% PURPOSE : binary star article, section "Discussion"
% DATE    : 23 Nov 2016
% -----------------------------------------------------------------------------

\section{Discussion}\label{sec:discussion}

The semi-analytical method presented in section \ref{sec:methods} relies on a numerical integration of the equations of motion to extract the fundamental frequencies of the perturbers. These numbers are then used in the analytical model to calculate the location of secular resonances. An analytical approach involving simple expressions for the giant planet's secular frequency would facilitate the study of a wide variety of systems, including newly detected ones, without resorting to numerical integrations.

% -------------------------------------

\subsection{Purely analytical methods}

Here we test various analytical methods to calculate the fundamental frequencies $g_{j}$. 

The Laplace-Lagrange model (LL) can also be used to calculate the secular frequencies of an arbitrary number $N$ of mutually interacting massive bodies \citep[see][section 7]{Mur1999}. The resulting frequencies will be reliable only if the assumptions of low eccentricity and inclination are fulfilled.

% --- Heppenheimer model ---
In the \citet{Hep1978} model (HEP) a restricted three-body problem is assumed, where the planet's mass is negligible relative to the masses of the two stars. The disturbing function is limited to terms of $O(e_{P}^{2})$ in the planet's eccentricity, but it can handle an arbitrary eccentricity $e_{B}$ of the binary \citep[see][]{And2016}. The expression for the forced secular frequency of the giant planet is
\begin{equation} \label{eq:heppenheimer}
  g_P = \frac{3}{4} \left( \frac{m_B}{m_A} \right) \left( \frac{a_P}{a_B} \right)^3 n_P (1 - e_B^2)^{-3/2},
\end{equation}
\edit{where $a_P$ ($n_P$) is the giant planet's semi-major axis (mean motion), $m_A$ and $m_B$ are the stellar masses, $a_B$ is the secondary star's distance from the primary, and $e_B$ its eccentricity.}

% --- Giuppone model ---
\citet{Giu2011} constructed a secular model (GIU) building on the Heppenheimer model, but extending it to second order in the masses. Instead of using the full analytical expressions they fitted a simple empirical formula for the secular frequency to their model, which is given by
\begin{equation} \label{eq:giuppone}
  g_P = g_0 \left[ 1 + 32 \left( \frac{m_B}{m_A} \right) \left( \frac{a_P}{a_B} \right)^2 (1 - e_B^2)^{-5} \right],
\end{equation}
where $g_0$ is given by the expression in equation \eq{eq:heppenheimer}. Strictly speaking, their formula was fitted to the $\gamma$ Cephei system, and was not intended to be used elsewhere.

% --- Georgakarakos model ---
In a series of papers, \citet{Geo2002, Geo2003, Geo2006, Geo2009} worked out a model (GEO) for hierarchical triple systems, i.e. a close binary star (or star-planet pair) with a much more distant third object. He derived closed formulas for the eccentricity evolution of \edit{the inner and outer object, as well as the} secular frequency going to second-order in the masses. In this model the eccentricity of the inner object should remain bounded below $\sim 0.2$, \edit{otherwise the neglected higher order terms in the eccentricity deteriorate the accuracy.} The expression for the secular frequency can be written as\footnote{\edit{Georgakarakos 2016, private communication}}
\begin{equation} \label{eq:georgakarakos}
  g_P = g_0 \left[ 1 + \frac{25}{8} \frac{m_B}{\sqrt{m_A (m_A + m_B)}} \left( \frac{a_P}{a_B} \right)^{3/2} \frac{3 + 2 e_{B}^{2}}{(1 - e_B^2)^{3/2}} \right],
\end{equation}
where again $g_0$ is given by equation \eq{eq:heppenheimer}.
Note that equation \eq{eq:georgakarakos} contains the complete analytical dependence on masses and semi-major axis ratio to the order of approximation, in contrast to equation \eq{eq:giuppone} which assumes an certain form and is different in important details.

\edit{The latter three models all assume a restricted three-body problem ($m_{P} = 0$). This choice mainly affects and limits the GEO model, which is also able to handle a massive planet (see details in \citealt{Geo2003}), but is of course less accurate when the planet's mass is neglected. We use the restricted problem primarily to permit for a fair comparison between the models.}
%, and hence they cannot be applied in case of more than one planet.

\begin{figure}
  \includegraphics[width=\columnwidth]{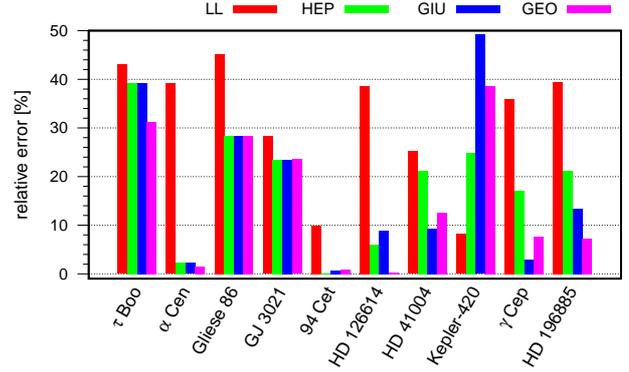}
  \caption{Magnitude of relative errors for the different analytical methods used to determine the fundamental secular frequency of the giant planet. The reference value for the frequency is derived from a numerical integration. Systems are ordered by increasing semi-major axis ratio $a_{P}/a_{B}$.}
  \label{fig:relerr}
\end{figure}

% --- describe figure 8 ---
\edit{Using these analytical methods (LL, HEP, GIU, GEO) we calculate the secular frequency of the giant planet, and compare the errors relative to the reference value obtained from a numerical simulation (without relativity) in \autoref{fig:relerr}.} In the HD~177830 system there are four massive bodies instead of three as presumed by the HEP, GIU, and GEO models; owing to this inherent inability we do not show HD~177830 in that figure. The LL method shows the largest overall errors, which is due to its construction of neglecting higher-order terms in the eccentricity. The one exception to this trend is the Kepler-420 case, here the LL performs best among the four methods. One reason for the mismatch in the analytical frequency for this particularly tight binary system is the combination of the semi-major axis ratio $\alpha = a_{P} / a_{B}$ (or orbital period ratio) with the mass of the secondary star. Comparing the other methods we see that they agree fairly well in most cases. \edit{For the largest separation binary 94~Ceti all errors are below 10~\%.}

\begin{table}
  \caption{Comparing the location of the secular resonance obtained by combinations of different analytical methods with the reference semi-analytical method. For each pair the first item indicates the model to calculate the giant planet's frequency, the second item the model for the test planet frequency. Abbreviations: LL = Laplace-Lagrange, HEP = Heppenheimer, GIU = Giuppone, GEO = Georgakarakos.}\label{tab:resloc}
  \begin{tabular}{ l *{4}{c} }
    \hline

Method          & \multicolumn{4}{c}{Resonance Location (au) for system} \\
                & HD 41004 & HD 126614 & HD 196885 & $\gamma$ Cep \\

    \hline

semi-analytical & 0.37     & 0.91      & 1.11      & 0.78 \\
LL/LL           & 0.30     & 0.70      & 0.87      & 0.62 \\
HEP/HEP         & 0.32     & 0.98      & 1.06      & 0.93 \\
HEP/LL          & 0.32     & 0.88      & 0.99      & 0.71 \\
GIU/GIU         & 0.35     & 1.08      & 1.29      & 1.06 \\
GIU/LL          & 0.35     & 0.94      & 1.17      & 0.79 \\
GEO/GEO         & 0.34     & 1.02      & 1.17      & 0.99 \\
GEO/LL          & 0.34     & 0.91      & 1.07      & 0.75 \\

    \hline
  \end{tabular}
\end{table}

% --- describe table 3 ---
\autoref{tab:resloc} compares the performance of the analytical methods for the exterior giant planet cases from \autoref{fig:periods}. The semi-analytical method serves as a reference (cf. locations in \autoref{tab:hzlim}), where we used a combination of a numerical integration of the respective system for the giant planet frequency, and the LL model for the test planet frequency. All other entries are pairs of methods, where the first item identifies the model for the giant planet's frequency, and the second item the model for the test planet frequencies.
\edit{Comparing mixed pairs involving the LL method, we see that usually they are closer to the numerical data than pairs of purely analytical methods. This behaviour can be understood when recalling that most importantly the LL method takes into account the contributions from the giant planet and secondary star (restricted 4-body problem), while the HEP, GIU, and GEO methods are limited to a restricted 3-body problem (giant planet only). Another issue might be the convergence rate of the Legendre expansions, see \citet{And2016} for a discussion.}

% -------------------------------------

\subsection{Additional improvements}

% --- second order models ---
A crucial ingredient is the giant planet's secular frequency $g_{P}$; differences in this frequency (cf. \autoref{fig:relerr}) transform into deviations $\Delta a$ of several 0.1~au in the location of the resonance. To improve on the analytical estimate of this frequency one needs second-order models in the masses including higher order terms in the planet's eccentricity.

This path has been followed recently by \citet{And2016}. The authors constructed a second-order model (in the masses) and determined the limits of its validity. The model complexity increased considerably as compared to the first order model, but they could show that for some of their investigated systems only the second order model was capable to correctly describe the dynamics. However, for a few extreme cases neither the first nor the second order model was sufficient, in those cases a direct numerical integration remains the only viable option. \citet{Lib2005} used a similar approach with a high order expansion of the disturbing function in the eccentricities to cope with moderately to highly eccentric extrasolar systems. \citet{Lib2013} extended the previous work by also moving to a second order theory with respect to the masses.

% --- new model of Nikos ---
\edit{Another approach is to stay with a first order model in the masses, but to use a high order expansion of the perturbing potential. \citet{Geo2016} derived an expansion up to order 11 in the semi-major axis ratio $\alpha$ valid for a terrestrial planet -- giant planet pair. In this way he obtained refined estimates for the secular frequency and maximum eccentricity of the smaller planet.}

% --- non-coplanar systems ---
So far we have considered only coplanar systems, but the orbital inclination is another important parameter. \citet{Pil2016} demonstrate for the HD~41004 system that the secular resonance can be traced numerically for inclinations up to $i \sim 40^{\circ}$. \citet{Wil1981} presented a method to map the secular resonance surfaces in the proper $(a,e,i)$ space. They applied this method to the solar system and investigated the main linear secular resonances (with Jupiter and Saturn) in the asteroid main belt.

% -----------------------------------------------------------------------------

% -----------------------------------------------------------------------------

% section "Conclusions"
% -----------------------------------------------------------------------------
% FILE    : concl.tex
% PURPOSE : binary star article, section "Conclusions"
% DATE    : 23 Nov 2016
% -----------------------------------------------------------------------------

\section{Summary}\label{sec:conclusions}

\edit{We investigated 11 binary star systems with detected planets in circumstellar (S-type) motion, and we calculated the location and extent of the habitable zone for these systems.} In most cases the presence of yet undetected Earth-mass planets cannot be ruled out. Using the semi-analytical method of \citet{Pil2016} we showed that in two systems (HD~177830 and $\alpha$~Cen) a \edit{linear SR is located directly in the HZ}. For three other binaries (HD~126614, HD~41004, and HD~196885) we found a SR close to the HZ. Our analysis revealed that giant planets located exterior to the HZ generally cause a linear SR, but not necessarily in the HZ. A moderate eccentricity of the giant planet ($e_{P} \ge 0.1$) is already sufficient to excite the forced eccentricities of terrestrial planets to values of $e \sim 1$ through the overlapping of the SR with nearby MMRs, where the secular forcing enhances the effect of the MMR. Therefore it is important to determine the location of the SR accurately, which underlines the usefulness of such studies. In the case of giant planets interior to the HZ we can exclude SR. However, for close-in giant planets ($a_{P} \le 0.1$~AU) the general relativistic precession of the pericenter can still lead to a SR, this is shown for two systems ($\tau$~Boo and $\alpha$~Cen). We compared the accuracy of our semi-analytical method to three analytical models, which can be used instead of a numerical integration to calculate the secular frequency of the giant planet. These analytical models are able to determine the resonance location to about $\Delta a \sim 0.1$~au, provided that the Laplace-Lagrange model is used for the test planet frequencies.

% -----------------------------------------------------------------------------

\section*{Acknowledgements}

The authors acknowledge support by FWF projects S11608-N16 and P22603-N16.
\edit{Our special thanks go to Nikolaos Georgakarakos for valuable discussions and helpful comments about the application of the analytical models. We also thank our referee for constructive comments that helped to improve the manuscript.}
For this investigation we made use of the Extrasolar Planet Encyclopaedia\footnote{\url{http://exoplanet.eu}} maintained by J. Schneider, and the Catalogue of Exoplanets in Binary Star Systems\footnote{\url{http://www.univie.ac.at/adg/schwarz/multiple.html}} maintained by R. Schwarz \citep{Sch2016}.

% -----------------------------------------------------------------------------

% -----------------------------------------------------------------------------

% ----- REFERENCES -----
\bibliographystyle{mnras}
\bibliography{literatur}

\begin{thebibliography}{}
\makeatletter
\relax
\def\mn@urlcharsother{\let\do\@makeother \do\$\do\&\do\#\do\^\do\_\do\%\do\~}
\def\mn@doi{\begingroup\mn@urlcharsother \@ifnextchar [ {\mn@doi@}
  {\mn@doi@[]}}
\def\mn@doi@[#1]#2{\def\@tempa{#1}\ifx\@tempa\@empty \href
  {http://dx.doi.org/#2} {doi:#2}\else \href {http://dx.doi.org/#2} {#1}\fi
  \endgroup}
\def\mn@eprint#1#2{\mn@eprint@#1:#2::\@nil}
\def\mn@eprint@arXiv#1{\href {http://arxiv.org/abs/#1} {{\tt arXiv:#1}}}
\def\mn@eprint@dblp#1{\href {http://dblp.uni-trier.de/rec/bibtex/#1.xml}
  {dblp:#1}}
\def\mn@eprint@#1:#2:#3:#4\@nil{\def\@tempa {#1}\def\@tempb {#2}\def\@tempc
  {#3}\ifx \@tempc \@empty \let \@tempc \@tempb \let \@tempb \@tempa \fi \ifx
  \@tempb \@empty \def\@tempb {arXiv}\fi \@ifundefined
  {mn@eprint@\@tempb}{\@tempb:\@tempc}{\expandafter \expandafter \csname
  mn@eprint@\@tempb\endcsname \expandafter{\@tempc}}}

\bibitem[\protect\citeauthoryear{{Andrade-Ines}, {Beaug{\'e}}, {Michtchenko}
  \& {Robutel}}{{Andrade-Ines} et~al.}{2016}]{And2016}
{Andrade-Ines} E.,  {Beaug{\'e}} C.,  {Michtchenko} T.,   {Robutel} P.,  2016,
  \mn@doi [CMDA] {10.1007/s10569-015-9669-5}, \href
  {http://esoads.eso.org/abs/2015CeMDA.tmp...72A} {124, 405}

\bibitem[\protect\citeauthoryear{{Bancelin}, {Hestroffer}  \&
  {Thuillot}}{{Bancelin} et~al.}{2012}]{Ban2012}
{Bancelin} D.,  {Hestroffer} D.,   {Thuillot} W.,  2012, \mn@doi [CMDA]
  {10.1007/s10569-011-9393-8}, \href
  {http://adsabs.harvard.edu/abs/2012CeMDA.112..221B} {112, 221}

\bibitem[\protect\citeauthoryear{{Bancelin}, {Pilat-Lohinger}, {Eggl},
  {Maindl}, {Sch{\"a}fer}, {Speith}  \& {Dvorak}}{{Bancelin}
  et~al.}{2015}]{Ban2015}
{Bancelin} D.,  {Pilat-Lohinger} E.,  {Eggl} S.,  {Maindl} T.~I.,
  {Sch{\"a}fer} C.,  {Speith} R.,   {Dvorak} R.,  2015, \mn@doi [\aap]
  {10.1051/0004-6361/201526430}, \href
  {http://cdsads.u-strasbg.fr/abs/2015A%26A...581A..46B} {581, A46}

\bibitem[\protect\citeauthoryear{{Bancelin}, {Pilat-Lohinger}  \&
  {Bazs{\'o}}}{{Bancelin} et~al.}{2016}]{Ban2016}
{Bancelin} D.,  {Pilat-Lohinger} E.,   {Bazs{\'o}} {\'A}.,  2016, \mn@doi
  [\aap] {10.1051/0004-6361/201528035}, \href
  {http://cdsads.u-strasbg.fr/abs/2016A%26A...591A.120B} {591, A120}

\bibitem[\protect\citeauthoryear{{Barclay}, {Quintana}, {Adams}, {Ciardi},
  {Huber}, {Foreman-Mackey}, {Montet}  \& {Caldwell}}{{Barclay}
  et~al.}{2015}]{Bar2015}
{Barclay} T.,  {Quintana} E.~V.,  {Adams} F.~C.,  {Ciardi} D.~R.,  {Huber} D.,
  {Foreman-Mackey} D.,  {Montet} B.~T.,   {Caldwell} D.,  2015, \mn@doi [\apj]
  {10.1088/0004-637X/809/1/7}, \href
  {http://adsabs.harvard.edu/abs/2015ApJ...809....7B} {809, 7}

\bibitem[\protect\citeauthoryear{{Beaug{\'e}} \& {Nesvorn{\'y}}}{{Beaug{\'e}}
  \& {Nesvorn{\'y}}}{2012}]{Bea2012b}
{Beaug{\'e}} C.,  {Nesvorn{\'y}} D.,  2012, \mn@doi [\apj]
  {10.1088/0004-637X/751/2/119}, \href
  {http://adsabs.harvard.edu/abs/2012ApJ...751..119B} {751, 119}

\bibitem[\protect\citeauthoryear{{Beaug{\'e}}, {Ferraz-Mello}  \&
  {Michtchenko}}{{Beaug{\'e}} et~al.}{2012}]{Bea2012a}
{Beaug{\'e}} C.,  {Ferraz-Mello} S.,   {Michtchenko} T.~A.,  2012, \mn@doi
  [Research in Astronomy and Astrophysics] {10.1088/1674-4527/12/8/009}, \href
  {http://adsabs.harvard.edu/abs/2012RAA....12.1044B} {12, 1044}

\bibitem[\protect\citeauthoryear{{Beutler}}{{Beutler}}{2005}]{Beu2005}
{Beutler} G.,  2005, {Methods of celestial mechanics. Vol. I: Physical,
  mathematical, and numerical principles}.
Astronomy and Astrophysics Library, Springer, Berlin

\bibitem[\protect\citeauthoryear{{Boss}}{{Boss}}{2006}]{Boss2006}
{Boss} A.~P.,  2006, \mn@doi [\apj] {10.1086/500530}, \href
  {http://adsabs.harvard.edu/abs/2006ApJ...641.1148B} {641, 1148}

\bibitem[\protect\citeauthoryear{{Brasser}, {Morbidelli}, {Gomes}, {Tsiganis}
  \& {Levison}}{{Brasser} et~al.}{2009}]{Bra2009}
{Brasser} R.,  {Morbidelli} A.,  {Gomes} R.,  {Tsiganis} K.,   {Levison} H.~F.,
   2009, \mn@doi [\aap] {10.1051/0004-6361/200912878}, \href
  {http://adsabs.harvard.edu/abs/2009A%26A...507.1053B} {507, 1053}

\bibitem[\protect\citeauthoryear{{Bromley} \& {Kenyon}}{{Bromley} \&
  {Kenyon}}{2015}]{Bro2015}
{Bromley} B.~C.,  {Kenyon} S.~J.,  2015, \mn@doi [\apj]
  {10.1088/0004-637X/806/1/98}, \href
  {http://adsabs.harvard.edu/abs/2015ApJ...806...98B} {806, 98}

\bibitem[\protect\citeauthoryear{{Brumberg}}{{Brumberg}}{2007}]{Bru2007}
{Brumberg} V.,  2007, \mn@doi [CMDA] {10.1007/s10569-007-9094-5}, \href
  {http://adsabs.harvard.edu/abs/2007CeMDA..99..245B} {99, 245}

\bibitem[\protect\citeauthoryear{{Butler}, {Marcy}, {Williams}, {Hauser}  \&
  {Shirts}}{{Butler} et~al.}{1997}]{But1997}
{Butler} R.~P.,  {Marcy} G.~W.,  {Williams} E.,  {Hauser} H.,   {Shirts} P.,
  1997, \mn@doi [\apjl] {10.1086/310444}, \href
  {http://adsabs.harvard.edu/abs/1997ApJ...474L.115B} {474, L115}

\bibitem[\protect\citeauthoryear{{Butler} et~al.,}{{Butler}
  et~al.}{2006}]{But2006}
{Butler} R.~P.,  et~al., 2006, \mn@doi [\apj] {10.1086/504701}, \href
  {http://esoads.eso.org/abs/2006ApJ...646..505B} {646, 505}

\bibitem[\protect\citeauthoryear{{Campante} et~al.,}{{Campante}
  et~al.}{2015}]{Cam2015}
{Campante} T.~L.,  et~al., 2015, \mn@doi [\apj] {10.1088/0004-637X/799/2/170},
  \href {http://adsabs.harvard.edu/abs/2015ApJ...799..170C} {799, 170}

\bibitem[\protect\citeauthoryear{{Chabrier}}{{Chabrier}}{2001}]{Cha2001}
{Chabrier} G.,  2001, \mn@doi [\apj] {10.1086/321401}, \href
  {http://adsabs.harvard.edu/abs/2001ApJ...554.1274C} {554, 1274}

\bibitem[\protect\citeauthoryear{{Chambers}}{{Chambers}}{1999}]{Cha1999}
{Chambers} J.~E.,  1999, \mn@doi [\mnras] {10.1046/j.1365-8711.1999.02379.x},
  \href {http://adsabs.harvard.edu/abs/1999MNRAS.304..793C} {304, 793}

\bibitem[\protect\citeauthoryear{{Chauvin}, {Lagrange}, {Udry}  \&
  {Mayor}}{{Chauvin} et~al.}{2007}]{Cha2007}
{Chauvin} G.,  {Lagrange} A.-M.,  {Udry} S.,   {Mayor} M.,  2007, \mn@doi
  [\aap] {10.1051/0004-6361:20067046}, \href
  {http://adsabs.harvard.edu/abs/2007A%26A...475..723C} {475, 723}

\bibitem[\protect\citeauthoryear{{Chauvin}, {Beust}, {Lagrange}  \&
  {Eggenberger}}{{Chauvin} et~al.}{2011}]{Cha2011}
{Chauvin} G.,  {Beust} H.,  {Lagrange} A.-M.,   {Eggenberger} A.,  2011,
  \mn@doi [\aap] {10.1051/0004-6361/201015433}, \href
  {http://adsabs.harvard.edu/abs/2011A%26A...528A...8C} {528, A8}

\bibitem[\protect\citeauthoryear{{Correia} et~al.,}{{Correia}
  et~al.}{2008}]{Cor2008}
{Correia} A.~C.~M.,  et~al., 2008, \mn@doi [\aap] {10.1051/0004-6361:20078908},
  \href {http://adsabs.harvard.edu/abs/2008A%26A...479..271C} {479, 271}

\bibitem[\protect\citeauthoryear{{Cuntz}}{{Cuntz}}{2014}]{Cun2014}
{Cuntz} M.,  2014, \mn@doi [\apj] {10.1088/0004-637X/780/1/14}, \href
  {http://adsabs.harvard.edu/abs/2014ApJ...780...14C} {780, 14}

\bibitem[\protect\citeauthoryear{{Desidera} \& {Barbieri}}{{Desidera} \&
  {Barbieri}}{2007}]{Des2007}
{Desidera} S.,  {Barbieri} M.,  2007, \mn@doi [\aap]
  {10.1051/0004-6361:20066319}, \href
  {http://esoads.eso.org/abs/2007A%26A...462..345D} {462, 345}

\bibitem[\protect\citeauthoryear{{Dodson-Robinson}, {Veras}, {Ford}  \&
  {Beichman}}{{Dodson-Robinson} et~al.}{2009}]{Dod2009}
{Dodson-Robinson} S.~E.,  {Veras} D.,  {Ford} E.~B.,   {Beichman} C.~A.,  2009,
  \mn@doi [\apj] {10.1088/0004-637X/707/1/79}, \href
  {http://adsabs.harvard.edu/abs/2009ApJ...707...79D} {707, 79}

\bibitem[\protect\citeauthoryear{{Duch{\^e}ne} \& {Kraus}}{{Duch{\^e}ne} \&
  {Kraus}}{2013}]{Duc2013}
{Duch{\^e}ne} G.,  {Kraus} A.,  2013, \mn@doi [\araa]
  {10.1146/annurev-astro-081710-102602}, \href
  {http://esoads.eso.org/abs/2013ARA%26A..51..269D} {51, 269}

\bibitem[\protect\citeauthoryear{{Dumusque} et~al.,}{{Dumusque}
  et~al.}{2012}]{Dum2012}
{Dumusque} X.,  et~al., 2012, \mn@doi [Nature] {10.1038/nature11572}, \href
  {http://adsabs.harvard.edu/abs/2012Natur.491..207D} {491, 207}

\bibitem[\protect\citeauthoryear{{Duquennoy} \& {Mayor}}{{Duquennoy} \&
  {Mayor}}{1991}]{Duq1991}
{Duquennoy} A.,  {Mayor} M.,  1991, \aap, \href
  {http://adsabs.harvard.edu/abs/1991A%26A...248..485D} {248, 485}

\bibitem[\protect\citeauthoryear{{Dvorak}, {Froeschle}  \&
  {Froeschle}}{{Dvorak} et~al.}{1989}]{Dvo1989}
{Dvorak} R.,  {Froeschle} C.,   {Froeschle} C.,  1989, \aap, \href
  {http://adsabs.harvard.edu/abs/1989A%26A...226..335D} {226, 335}

\bibitem[\protect\citeauthoryear{{Eggl} \& {Dvorak}}{{Eggl} \&
  {Dvorak}}{2010}]{Eggl2010}
{Eggl} S.,  {Dvorak} R.,  2010, in {Souchay} J.,  {Dvorak} R.,  eds,  Lecture
  Notes in Physics, Berlin Springer Verlag Vol. 790, Dynamics of Small Solar
  System Bodies and Exoplanets. pp 431--480,
  \mn@doi{10.1007/978-3-642-04458-8_9}

\bibitem[\protect\citeauthoryear{{Eggl}, {Pilat-Lohinger}, {Georgakarakos},
  {Gyergyovits}  \& {Funk}}{{Eggl} et~al.}{2012}]{Eggl2012}
{Eggl} S.,  {Pilat-Lohinger} E.,  {Georgakarakos} N.,  {Gyergyovits} M.,
  {Funk} B.,  2012, \mn@doi [\apj] {10.1088/0004-637X/752/1/74}, \href
  {http://adsabs.harvard.edu/abs/2012ApJ...752...74E} {752, 74}

\bibitem[\protect\citeauthoryear{{Eggl}, {Haghighipour}  \&
  {Pilat-Lohinger}}{{Eggl} et~al.}{2013}]{Eggl2013}
{Eggl} S.,  {Haghighipour} N.,   {Pilat-Lohinger} E.,  2013, \mn@doi [\apj]
  {10.1088/0004-637X/764/2/130}, \href
  {http://adsabs.harvard.edu/abs/2013ApJ...764..130E} {764, 130}

\bibitem[\protect\citeauthoryear{{Endl}, {Cochran}, {Hatzes}  \&
  {Wittenmyer}}{{Endl} et~al.}{2011}]{Endl2011}
{Endl} M.,  {Cochran} W.~D.,  {Hatzes} A.~P.,   {Wittenmyer} R.~A.,  2011, in
  {Schuh} S.,  {Drechsel} H.,   {Heber} U.,  eds,  American Institute of
  Physics Conference Series Vol. 1331, American Institute of Physics Conference
  Series. pp 88--94 (\mn@eprint {arXiv} {1101.2588}),
  \mn@doi{10.1063/1.3556187}

\bibitem[\protect\citeauthoryear{{Endl} et~al.,}{{Endl}
  et~al.}{2015}]{Endl2015}
{Endl} M.,  et~al., 2015, \mn@doi [International Journal of Astrobiology]
  {10.1017/S1473550414000081}, \href
  {http://adsabs.harvard.edu/abs/2015IJAsB..14..305E} {14, 305}

\bibitem[\protect\citeauthoryear{{Everhart}}{{Everhart}}{1974}]{Eve1974}
{Everhart} E.,  1974, \mn@doi [Cel. Mech.] {10.1007/BF01261877}, \href
  {http://esoads.eso.org/abs/1974CeMec..10...35E} {10, 35}

\bibitem[\protect\citeauthoryear{{Fabrycky} \& {Tremaine}}{{Fabrycky} \&
  {Tremaine}}{2007}]{Fab2007}
{Fabrycky} D.,  {Tremaine} S.,  2007, \mn@doi [\apj] {10.1086/521702}, \href
  {http://adsabs.harvard.edu/abs/2007ApJ...669.1298F} {669, 1298}

\bibitem[\protect\citeauthoryear{{Fischer} et~al.,}{{Fischer}
  et~al.}{2016}]{Fis2016}
{Fischer} D.~A.,  et~al., 2016, \mn@doi [\pasp]
  {10.1088/1538-3873/128/964/066001}, \href
  {http://adsabs.harvard.edu/abs/2016PASP..128f6001F} {128, 066001}

\bibitem[\protect\citeauthoryear{{Fogg} \& {Nelson}}{{Fogg} \&
  {Nelson}}{2007}]{Fogg2007}
{Fogg} M.~J.,  {Nelson} R.~P.,  2007, \mn@doi [\aap]
  {10.1051/0004-6361:20077950}, \href
  {http://adsabs.harvard.edu/abs/2007A%26A...472.1003F} {472, 1003}

\bibitem[\protect\citeauthoryear{{Forgan}}{{Forgan}}{2016}]{For2016}
{Forgan} D.,  2016, \mn@doi [\mnras] {10.1093/mnras/stw2098}, \href
  {http://adsabs.harvard.edu/abs/2016MNRAS.463.2768F} {463, 2768}

\bibitem[\protect\citeauthoryear{Frigo \& Johnson}{Frigo \&
  Johnson}{2005}]{Fri2005}
Frigo M.,  Johnson S.~G.,  2005, Proceedings of the IEEE, 93, 216

\bibitem[\protect\citeauthoryear{{Froeschl{\'e}}, {Lega}  \&
  {Gonczi}}{{Froeschl{\'e}} et~al.}{1997}]{Fro1997}
{Froeschl{\'e}} C.,  {Lega} E.,   {Gonczi} R.,  1997, \mn@doi [CMDA]
  {10.1023/A:1008276418601}, \href
  {http://adsabs.harvard.edu/abs/1997CeMDA..67...41F} {67, 41}

\bibitem[\protect\citeauthoryear{{Fuhrmann}, {Chini}, {Buda}  \& {Pozo
  Nu{\~n}ez}}{{Fuhrmann} et~al.}{2014}]{Fuh2014}
{Fuhrmann} K.,  {Chini} R.,  {Buda} L.-S.,   {Pozo Nu{\~n}ez} F.,  2014,
  \mn@doi [\apj] {10.1088/0004-637X/785/1/68}, \href
  {http://adsabs.harvard.edu/abs/2014ApJ...785...68F} {785, 68}

\bibitem[\protect\citeauthoryear{{Funk}, {Pilat-Lohinger}  \& {Eggl}}{{Funk}
  et~al.}{2015}]{Funk2015}
{Funk} B.,  {Pilat-Lohinger} E.,   {Eggl} S.,  2015, \mn@doi [\mnras]
  {10.1093/mnras/stv253}, \href
  {http://adsabs.harvard.edu/abs/2015MNRAS.448.3797F} {448, 3797}

\bibitem[\protect\citeauthoryear{{Georgakarakos}}{{Georgakarakos}}{2002}]{Geo2002}
{Georgakarakos} N.,  2002, \mn@doi [\mnras] {10.1046/j.1365-8711.2002.05936.x},
  \href {http://esoads.eso.org/abs/2002MNRAS.337..559G} {337, 559}

\bibitem[\protect\citeauthoryear{{Georgakarakos}}{{Georgakarakos}}{2003}]{Geo2003}
{Georgakarakos} N.,  2003, \mn@doi [\mnras] {10.1046/j.1365-8711.2003.06942.x},
  \href {http://adsabs.harvard.edu/abs/2003MNRAS.345..340G} {345, 340}

\bibitem[\protect\citeauthoryear{{Georgakarakos}}{{Georgakarakos}}{2006}]{Geo2006}
{Georgakarakos} N.,  2006, \mn@doi [\mnras] {10.1111/j.1365-2966.2005.09888.x},
  \href {http://adsabs.harvard.edu/abs/2006MNRAS.366..566G} {366, 566}

\bibitem[\protect\citeauthoryear{{Georgakarakos}}{{Georgakarakos}}{2009}]{Geo2009}
{Georgakarakos} N.,  2009, \mn@doi [\mnras] {10.1111/j.1365-2966.2008.14143.x},
  \href {http://adsabs.harvard.edu/abs/2009MNRAS.392.1253G} {392, 1253}

\bibitem[\protect\citeauthoryear{{Georgakarakos}, {Dobbs-Dixon}  \&
  {Way}}{{Georgakarakos} et~al.}{2016}]{Geo2016}
{Georgakarakos} N.,  {Dobbs-Dixon} I.,   {Way} M.~J.,  2016, \mn@doi [\mnras]
  {10.1093/mnras/stw1378}, \href
  {http://adsabs.harvard.edu/abs/2016MNRAS.461.1512G} {461, 1512}

\bibitem[\protect\citeauthoryear{{Giuppone}, {Leiva}, {Correa-Otto}  \&
  {Beaug{\'e}}}{{Giuppone} et~al.}{2011}]{Giu2011}
{Giuppone} C.~A.,  {Leiva} A.~M.,  {Correa-Otto} J.,   {Beaug{\'e}} C.,  2011,
  \mn@doi [\aap] {10.1051/0004-6361/201016375}, \href
  {http://adsabs.harvard.edu/abs/2011A%26A...530A.103G} {530, A103}

\bibitem[\protect\citeauthoryear{{Gould} et~al.,}{{Gould}
  et~al.}{2014}]{Gou2014}
{Gould} A.,  et~al., 2014, \mn@doi [Science] {10.1126/science.1251527}, \href
  {http://adsabs.harvard.edu/abs/2014Sci...345...46G} {345, 46}

\bibitem[\protect\citeauthoryear{{Haghighipour} \& {Raymond}}{{Haghighipour} \&
  {Raymond}}{2007}]{Hag2007}
{Haghighipour} N.,  {Raymond} S.~N.,  2007, \mn@doi [\apj] {10.1086/520501},
  \href {http://adsabs.harvard.edu/abs/2007ApJ...666..436H} {666, 436}

\bibitem[\protect\citeauthoryear{{Hanslmeier} \& {Dvorak}}{{Hanslmeier} \&
  {Dvorak}}{1984}]{Han1984}
{Hanslmeier} A.,  {Dvorak} R.,  1984, \aap, \href
  {http://adsabs.harvard.edu/abs/1984A%26A...132..203H} {132, 203}

\bibitem[\protect\citeauthoryear{{Hatzes}}{{Hatzes}}{2013}]{Hat2013}
{Hatzes} A.~P.,  2013, \mn@doi [\apj] {10.1088/0004-637X/770/2/133}, \href
  {http://adsabs.harvard.edu/abs/2013ApJ...770..133H} {770, 133}

\bibitem[\protect\citeauthoryear{{Hatzes}, {Cochran}, {Endl}, {McArthur},
  {Paulson}, {Walker}, {Campbell}  \& {Yang}}{{Hatzes} et~al.}{2003}]{Hat2003}
{Hatzes} A.~P.,  {Cochran} W.~D.,  {Endl} M.,  {McArthur} B.,  {Paulson} D.~B.,
   {Walker} G.~A.~H.,  {Campbell} B.,   {Yang} S.,  2003, \mn@doi [\apj]
  {10.1086/379281}, \href {http://adsabs.harvard.edu/abs/2003ApJ...599.1383H}
  {599, 1383}

\bibitem[\protect\citeauthoryear{{Heppenheimer}}{{Heppenheimer}}{1978}]{Hep1978}
{Heppenheimer} T.~A.,  1978, \aap, \href
  {http://adsabs.harvard.edu/abs/1978A%26A....65..421H} {65, 421}

\bibitem[\protect\citeauthoryear{{Holman} \& {Wiegert}}{{Holman} \&
  {Wiegert}}{1999}]{Hol1999}
{Holman} M.~J.,  {Wiegert} P.~A.,  1999, \mn@doi [AJ] {10.1086/300695}, \href
  {http://adsabs.harvard.edu/abs/1999AJ....117..621H} {117, 621}

\bibitem[\protect\citeauthoryear{{Howard} et~al.,}{{Howard}
  et~al.}{2010}]{How2010}
{Howard} A.~W.,  et~al., 2010, \mn@doi [\apj] {10.1088/0004-637X/721/2/1467},
  \href {http://adsabs.harvard.edu/abs/2010ApJ...721.1467H} {721, 1467}

\bibitem[\protect\citeauthoryear{{Jang-Condell}}{{Jang-Condell}}{2015}]{Jan2015}
{Jang-Condell} H.,  2015, \mn@doi [\apj] {10.1088/0004-637X/799/2/147}, \href
  {http://adsabs.harvard.edu/abs/2015ApJ...799..147J} {799, 147}

\bibitem[\protect\citeauthoryear{{Jofr{\'e}}, {Petrucci}, {Saffe}, {Saker}, {de
  la Villarmois}, {Chavero}, {G{\'o}mez}  \& {Mauas}}{{Jofr{\'e}}
  et~al.}{2015}]{Jof2015}
{Jofr{\'e}} E.,  {Petrucci} R.,  {Saffe} C.,  {Saker} L.,  {de la Villarmois}
  E.~A.,  {Chavero} C.,  {G{\'o}mez} M.,   {Mauas} P.~J.~D.,  2015, \mn@doi
  [\aap] {10.1051/0004-6361/201424474}, \href
  {http://cdsads.u-strasbg.fr/abs/2015A%26A...574A..50J} {574, A50}

\bibitem[\protect\citeauthoryear{{Kaltenegger} \& {Haghighipour}}{{Kaltenegger}
  \& {Haghighipour}}{2013}]{Kal2013}
{Kaltenegger} L.,  {Haghighipour} N.,  2013, \mn@doi [\apj]
  {10.1088/0004-637X/777/2/165}, \href
  {http://adsabs.harvard.edu/abs/2013ApJ...777..165K} {777, 165}

\bibitem[\protect\citeauthoryear{{Kasting}, {Whitmire}  \&
  {Reynolds}}{{Kasting} et~al.}{1993}]{Kas1993}
{Kasting} J.~F.,  {Whitmire} D.~P.,   {Reynolds} R.~T.,  1993, \mn@doi [Icarus]
  {10.1006/icar.1993.1010}, \href
  {http://adsabs.harvard.edu/abs/1993Icar..101..108K} {101, 108}

\bibitem[\protect\citeauthoryear{{Kley} \& {Nelson}}{{Kley} \&
  {Nelson}}{2010}]{Kley2010}
{Kley} W.,  {Nelson} R.~P.,  2010, in {Haghighipour} N.,  ed.,  Astrophysics
  and Space Science Library Vol. 366, Planets in Binary Star Systems. p.~135,
  \mn@doi{10.1007/978-90-481-8687-7_6}

\bibitem[\protect\citeauthoryear{{Knezevic}, {Milani}, {Farinella}, {Froeschle}
   \& {Froeschle}}{{Knezevic} et~al.}{1991}]{Kne1991}
{Knezevic} Z.,  {Milani} A.,  {Farinella} P.,  {Froeschle} C.,   {Froeschle}
  C.,  1991, \mn@doi [Icarus] {10.1016/0019-1035(91)90215-F}, \href
  {http://esoads.eso.org/abs/1991Icar...93..316K} {93, 316}

\bibitem[\protect\citeauthoryear{{Kopparapu} et~al.,}{{Kopparapu}
  et~al.}{2013}]{Kop2013}
{Kopparapu} R.~K.,  et~al., 2013, \mn@doi [\apj] {10.1088/0004-637X/765/2/131},
  \href {http://adsabs.harvard.edu/abs/2013ApJ...765..131K} {765, 131}

\bibitem[\protect\citeauthoryear{{Laskar}}{{Laskar}}{2008}]{Las2008}
{Laskar} J.,  2008, \mn@doi [Icarus] {10.1016/j.icarus.2008.02.017}, \href
  {http://adsabs.harvard.edu/abs/2008Icar..196....1L} {196, 1}

\bibitem[\protect\citeauthoryear{{Levison} \& {Agnor}}{{Levison} \&
  {Agnor}}{2003}]{Lev2003}
{Levison} H.~F.,  {Agnor} C.,  2003, \mn@doi [\aj] {10.1086/374625}, \href
  {http://esoads.eso.org/abs/2003AJ....125.2692L} {125, 2692}

\bibitem[\protect\citeauthoryear{{Libert} \& {Henrard}}{{Libert} \&
  {Henrard}}{2005}]{Lib2005}
{Libert} A.-S.,  {Henrard} J.,  2005, \mn@doi [CMDA]
  {10.1007/s10569-005-0181-1}, \href
  {http://esoads.eso.org/abs/2005CeMDA..93..187L} {93, 187}

\bibitem[\protect\citeauthoryear{{Libert} \& {Sansottera}}{{Libert} \&
  {Sansottera}}{2013}]{Lib2013}
{Libert} A.-S.,  {Sansottera} M.,  2013, \mn@doi [CMDA]
  {10.1007/s10569-013-9501-z}, \href
  {http://esoads.eso.org/abs/2013CeMDA.117..149L} {117, 149}

\bibitem[\protect\citeauthoryear{{Liebert}, {Bergeron}  \& {Holberg}}{{Liebert}
  et~al.}{2005}]{Lie2005}
{Liebert} J.,  {Bergeron} P.,   {Holberg} J.~B.,  2005, \mn@doi [\apjs]
  {10.1086/425738}, \href {http://adsabs.harvard.edu/abs/2005ApJS..156...47L}
  {156, 47}

\bibitem[\protect\citeauthoryear{{Marois}, {Macintosh}, {Barman}, {Zuckerman},
  {Song}, {Patience}, {Lafreni{\`e}re}  \& {Doyon}}{{Marois}
  et~al.}{2008}]{Mar2008}
{Marois} C.,  {Macintosh} B.,  {Barman} T.,  {Zuckerman} B.,  {Song} I.,
  {Patience} J.,  {Lafreni{\`e}re} D.,   {Doyon} R.,  2008, \mn@doi [Science]
  {10.1126/science.1166585}, \href
  {http://adsabs.harvard.edu/abs/2008Sci...322.1348M} {322, 1348}

\bibitem[\protect\citeauthoryear{{Mayor}, {Udry}, {Naef}, {Pepe}, {Queloz},
  {Santos}  \& {Burnet}}{{Mayor} et~al.}{2004}]{May2004}
{Mayor} M.,  {Udry} S.,  {Naef} D.,  {Pepe} F.,  {Queloz} D.,  {Santos} N.~C.,
   {Burnet} M.,  2004, \mn@doi [\aap] {10.1051/0004-6361:20034250}, \href
  {http://adsabs.harvard.edu/abs/2004A%26A...415..391M} {415, 391}

\bibitem[\protect\citeauthoryear{{Mudryk} \& {Wu}}{{Mudryk} \&
  {Wu}}{2006}]{Mud2006}
{Mudryk} L.~R.,  {Wu} Y.,  2006, \mn@doi [\apj] {10.1086/499347}, \href
  {http://esoads.eso.org/abs/2006ApJ...639..423M} {639, 423}

\bibitem[\protect\citeauthoryear{{Mugrauer} \& {Neuh{\"a}user}}{{Mugrauer} \&
  {Neuh{\"a}user}}{2009}]{Mug2009}
{Mugrauer} M.,  {Neuh{\"a}user} R.,  2009, \mn@doi [\aap]
  {10.1051/0004-6361:200810639}, \href
  {http://adsabs.harvard.edu/abs/2009A%26A...494..373M} {494, 373}

\bibitem[\protect\citeauthoryear{{M{\"u}ller} \& {Kley}}{{M{\"u}ller} \&
  {Kley}}{2012}]{Mul2012}
{M{\"u}ller} T.~W.~A.,  {Kley} W.,  2012, \mn@doi [\aap]
  {10.1051/0004-6361/201118202}, \href
  {http://esoads.eso.org/abs/2012A%26A...539A..18M} {539, A18}

\bibitem[\protect\citeauthoryear{{Murray} \& {Dermott}}{{Murray} \&
  {Dermott}}{1999}]{Mur1999}
{Murray} C.~D.,  {Dermott} S.~F.,  1999, {Solar system dynamics}.
Cambridge University Press

\bibitem[\protect\citeauthoryear{{Naef}, {Mayor}, {Pepe}, {Queloz}, {Santos},
  {Udry}  \& {Burnet}}{{Naef} et~al.}{2001}]{Naef2001}
{Naef} D.,  {Mayor} M.,  {Pepe} F.,  {Queloz} D.,  {Santos} N.~C.,  {Udry} S.,
   {Burnet} M.,  2001, \mn@doi [\aap] {10.1051/0004-6361:20010841}, \href
  {http://adsabs.harvard.edu/abs/2001A%26A...375..205N} {375, 205}

\bibitem[\protect\citeauthoryear{{Naoz}, {Farr}  \& {Rasio}}{{Naoz}
  et~al.}{2012}]{Naoz2012}
{Naoz} S.,  {Farr} W.~M.,   {Rasio} F.~A.,  2012, \mn@doi [\apjl]
  {10.1088/2041-8205/754/2/L36}, \href
  {http://adsabs.harvard.edu/abs/2012ApJ...754L..36N} {754, L36}

\bibitem[\protect\citeauthoryear{{Ogihara}, {Kobayashi}  \&
  {Inutsuka}}{{Ogihara} et~al.}{2014}]{Ogi2014}
{Ogihara} M.,  {Kobayashi} H.,   {Inutsuka} S.-i.,  2014, \mn@doi [\apj]
  {10.1088/0004-637X/787/2/172}, \href
  {http://adsabs.harvard.edu/abs/2014ApJ...787..172O} {787, 172}

\bibitem[\protect\citeauthoryear{{Pilat-Lohinger} \& {Dvorak}}{{Pilat-Lohinger}
  \& {Dvorak}}{2002}]{Pil2002}
{Pilat-Lohinger} E.,  {Dvorak} R.,  2002, \mn@doi [CMDA]
  {10.1023/A:1014586308539}, \href
  {http://adsabs.harvard.edu/abs/2002CeMDA..82..143P} {82, 143}

\bibitem[\protect\citeauthoryear{{Pilat-Lohinger} \& {Funk}}{{Pilat-Lohinger}
  \& {Funk}}{2010}]{Pil2010}
{Pilat-Lohinger} E.,  {Funk} B.,  2010, in {Souchay} J.,  {Dvorak} R.,  eds,
  Lecture Notes in Physics, Berlin Springer Verlag Vol. 790, Dynamics of Small
  Solar System Bodies and Exoplanets. pp 481--510,
  \mn@doi{10.1007/978-3-642-04458-8_10}

\bibitem[\protect\citeauthoryear{{Pilat-Lohinger}, {Bazs{\'o}}  \&
  {Funk}}{{Pilat-Lohinger} et~al.}{2016}]{Pil2016}
{Pilat-Lohinger} E.,  {Bazs{\'o}} {\'A}.,   {Funk} B.,  2016, \mn@doi [\aj]
  {10.3847/0004-6256/152/5/139}, \href
  {http://esoads.eso.org/abs/2016AJ....152..139P} {152, 139}

\bibitem[\protect\citeauthoryear{{Poleski} et~al.,}{{Poleski}
  et~al.}{2014}]{Pol2014}
{Poleski} R.,  et~al., 2014, \mn@doi [\apj] {10.1088/0004-637X/795/1/42}, \href
  {http://adsabs.harvard.edu/abs/2014ApJ...795...42P} {795, 42}

\bibitem[\protect\citeauthoryear{{Queloz} et~al.,}{{Queloz}
  et~al.}{2000}]{Que2000}
{Queloz} D.,  et~al., 2000, \aap, \href
  {http://adsabs.harvard.edu/abs/2000A%26A...354...99Q} {354, 99}

\bibitem[\protect\citeauthoryear{{Quintana}, {Adams}, {Lissauer}  \&
  {Chambers}}{{Quintana} et~al.}{2007}]{Qui2007}
{Quintana} E.~V.,  {Adams} F.~C.,  {Lissauer} J.~J.,   {Chambers} J.~E.,  2007,
  \mn@doi [\apj] {10.1086/512542}, \href
  {http://adsabs.harvard.edu/abs/2007ApJ...660..807Q} {660, 807}

\bibitem[\protect\citeauthoryear{{Rabl} \& {Dvorak}}{{Rabl} \&
  {Dvorak}}{1988}]{Rabl1988}
{Rabl} G.,  {Dvorak} R.,  1988, \aap, \href
  {http://adsabs.harvard.edu/abs/1988A%26A...191..385R} {191, 385}

\bibitem[\protect\citeauthoryear{{Raghavan}, {Henry}, {Mason}, {Subasavage},
  {Jao}, {Beaulieu}  \& {Hambly}}{{Raghavan} et~al.}{2006}]{Rag2006}
{Raghavan} D.,  {Henry} T.~J.,  {Mason} B.~D.,  {Subasavage} J.~P.,  {Jao}
  W.-C.,  {Beaulieu} T.~D.,   {Hambly} N.~C.,  2006, \mn@doi [\apj]
  {10.1086/504823}, \href {http://adsabs.harvard.edu/abs/2006ApJ...646..523R}
  {646, 523}

\bibitem[\protect\citeauthoryear{{Raghavan} et~al.,}{{Raghavan}
  et~al.}{2010}]{Rag2010}
{Raghavan} D.,  et~al., 2010, \mn@doi [\apjs] {10.1088/0067-0049/190/1/1},
  \href {http://adsabs.harvard.edu/abs/2010ApJS..190....1R} {190, 1}

\bibitem[\protect\citeauthoryear{{Rajpaul}, {Aigrain}  \& {Roberts}}{{Rajpaul}
  et~al.}{2016}]{Raj2016}
{Rajpaul} V.,  {Aigrain} S.,   {Roberts} S.,  2016, \mn@doi [\mnras]
  {10.1093/mnrasl/slv164}, \href
  {http://adsabs.harvard.edu/abs/2016MNRAS.456L...6R} {456, L6}

\bibitem[\protect\citeauthoryear{{Ramirez} \& {Kaltenegger}}{{Ramirez} \&
  {Kaltenegger}}{2016}]{Ram2016}
{Ramirez} R.~M.,  {Kaltenegger} L.,  2016, \mn@doi [\apj]
  {10.3847/0004-637X/823/1/6}, \href
  {http://esoads.eso.org/abs/2016ApJ...823....6R} {823, 6}

\bibitem[\protect\citeauthoryear{{Raymond}, {Quinn}  \& {Lunine}}{{Raymond}
  et~al.}{2005}]{Ray2005}
{Raymond} S.~N.,  {Quinn} T.,   {Lunine} J.~I.,  2005, \mn@doi [\icarus]
  {10.1016/j.icarus.2005.03.008}, \href
  {http://adsabs.harvard.edu/abs/2005Icar..177..256R} {177, 256}

\bibitem[\protect\citeauthoryear{{Reegen}}{{Reegen}}{2007}]{Ree2007}
{Reegen} P.,  2007, \mn@doi [\aap] {10.1051/0004-6361:20066597}, \href
  {http://adsabs.harvard.edu/abs/2007A%26A...467.1353R} {467, 1353}

\bibitem[\protect\citeauthoryear{{Roberts} Jr. et~al.,}{{Roberts}
  et~al.}{2015}]{Rob2015}
{Roberts} Jr. L.~C.,  et~al., 2015, \mn@doi [\aj]
  {10.1088/0004-6256/150/4/103}, \href
  {http://esoads.eso.org/abs/2015AJ....150..103R} {150, 103}

\bibitem[\protect\citeauthoryear{{Rodler}, {Lopez-Morales}  \&
  {Ribas}}{{Rodler} et~al.}{2012}]{Rod2012}
{Rodler} F.,  {Lopez-Morales} M.,   {Ribas} I.,  2012, \mn@doi [\apjl]
  {10.1088/2041-8205/753/1/L25}, \href
  {http://esoads.eso.org/abs/2012ApJ...753L..25R} {753, L25}

\bibitem[\protect\citeauthoryear{{Roell}, {Neuh{\"a}user}, {Seifahrt}  \&
  {Mugrauer}}{{Roell} et~al.}{2012}]{Roe2012}
{Roell} T.,  {Neuh{\"a}user} R.,  {Seifahrt} A.,   {Mugrauer} M.,  2012,
  \mn@doi [\aap] {10.1051/0004-6361/201118051}, \href
  {http://adsabs.harvard.edu/abs/2012A%26A...542A..92R} {542, A92}

\bibitem[\protect\citeauthoryear{{Rowe} et~al.,}{{Rowe}
  et~al.}{2014}]{Rowe2014}
{Rowe} J.~F.,  et~al., 2014, \mn@doi [\apj] {10.1088/0004-637X/784/1/45}, \href
  {http://adsabs.harvard.edu/abs/2014ApJ...784...45R} {784, 45}

\bibitem[\protect\citeauthoryear{{Santerne} et~al.,}{{Santerne}
  et~al.}{2014}]{San2014}
{Santerne} A.,  et~al., 2014, \mn@doi [\aap] {10.1051/0004-6361/201424158},
  \href {http://adsabs.harvard.edu/abs/2014A%26A...571A..37S} {571, A37}

\bibitem[\protect\citeauthoryear{{Schwarz}, {Funk}, {Zechner}  \&
  {Bazs{\'o}}}{{Schwarz} et~al.}{2016}]{Sch2016}
{Schwarz} R.,  {Funk} B.,  {Zechner} R.,   {Bazs{\'o}} {\'A}.,  2016, \mn@doi
  [\mnras] {10.1093/mnras/stw1218}, \href
  {http://adsabs.harvard.edu/abs/2016MNRAS.460.3598S} {460, 3598}

\bibitem[\protect\citeauthoryear{{Soubiran}, {Le Campion}, {Brouillet}  \&
  {Chemin}}{{Soubiran} et~al.}{2016}]{Sou2016}
{Soubiran} C.,  {Le Campion} J.-F.,  {Brouillet} N.,   {Chemin} L.,  2016,
  \mn@doi [\aap] {10.1051/0004-6361/201628497}, \href
  {http://adsabs.harvard.edu/abs/2016A%26A...591A.118S} {591, A118}

\bibitem[\protect\citeauthoryear{{Thebault} \& {Haghighipour}}{{Thebault} \&
  {Haghighipour}}{2015}]{The2015}
{Thebault} P.,  {Haghighipour} N.,  2015, in {Jin} S.,  {Haghighipour} N.,
  {Ip} W.-H.,  eds, Planetary Exploration and Science: Recent Results and
  Advances. Springer Geophysics, Springer-Verlag, Berlin Heidelberg.
pp 309--340 (\mn@eprint {arXiv} {1406.1357}),
  \mn@doi{10.1007/978-3-662-45052-9}

\bibitem[\protect\citeauthoryear{{Tokovinin}}{{Tokovinin}}{2014}]{Tok2014}
{Tokovinin} A.,  2014, \mn@doi [AJ] {10.1088/0004-6256/147/4/87}, \href
  {http://adsabs.harvard.edu/abs/2014AJ....147...87T} {147, 87}

\bibitem[\protect\citeauthoryear{{Torres}}{{Torres}}{2007}]{Tor2007}
{Torres} G.,  2007, \mn@doi [\apj] {10.1086/509715}, \href
  {http://esoads.eso.org/abs/2007ApJ...654.1095T} {654, 1095}

\bibitem[\protect\citeauthoryear{{Vogt}, {Marcy}, {Butler}  \& {Apps}}{{Vogt}
  et~al.}{2000}]{Vogt2000}
{Vogt} S.~S.,  {Marcy} G.~W.,  {Butler} R.~P.,   {Apps} K.,  2000, \mn@doi
  [\apj] {10.1086/308981}, \href
  {http://adsabs.harvard.edu/abs/2000ApJ...536..902V} {536, 902}

\bibitem[\protect\citeauthoryear{{Vorobyov}}{{Vorobyov}}{2013}]{Vor2013}
{Vorobyov} E.~I.,  2013, \mn@doi [\aap] {10.1051/0004-6361/201220601}, \href
  {http://adsabs.harvard.edu/abs/2013A%26A...552A.129V} {552, A129}

\bibitem[\protect\citeauthoryear{{Wang}, {Fischer}, {Xie}  \& {Ciardi}}{{Wang}
  et~al.}{2014}]{Wang2014}
{Wang} J.,  {Fischer} D.~A.,  {Xie} J.-W.,   {Ciardi} D.~R.,  2014, \mn@doi
  [\apj] {10.1088/0004-637X/791/2/111}, \href
  {http://esoads.eso.org/abs/2014ApJ...791..111W} {791, 111}

\bibitem[\protect\citeauthoryear{{Williams} \& {Faulkner}}{{Williams} \&
  {Faulkner}}{1981}]{Wil1981}
{Williams} J.~G.,  {Faulkner} J.,  1981, \mn@doi [Icarus]
  {10.1016/0019-1035(81)90140-8}, \href
  {http://esoads.eso.org/abs/1981Icar...46..390W} {46, 390}

\bibitem[\protect\citeauthoryear{{Williams} \& {Pollard}}{{Williams} \&
  {Pollard}}{2002}]{Wil2002}
{Williams} D.~M.,  {Pollard} D.,  2002, \mn@doi [International Journal of
  Astrobiology] {10.1017/S1473550402001064}, \href
  {http://adsabs.harvard.edu/abs/2002IJAsB...1...61W} {1, 61}

\bibitem[\protect\citeauthoryear{{Winn} \& {Fabrycky}}{{Winn} \&
  {Fabrycky}}{2015}]{Winn2015}
{Winn} J.~N.,  {Fabrycky} D.~C.,  2015, \mn@doi [\araa]
  {10.1146/annurev-astro-082214-122246}, \href
  {http://esoads.eso.org/abs/2015ARA%26A..53..409W} {53, 409}

\bibitem[\protect\citeauthoryear{{Wisdom}}{{Wisdom}}{1980}]{Wis1980}
{Wisdom} J.,  1980, \mn@doi [AJ] {10.1086/112778}, \href
  {http://adsabs.harvard.edu/abs/1980AJ.....85.1122W} {85, 1122}

\bibitem[\protect\citeauthoryear{{Zucker}, {Mazeh}, {Santos}, {Udry}  \&
  {Mayor}}{{Zucker} et~al.}{2004}]{Zuc2004}
{Zucker} S.,  {Mazeh} T.,  {Santos} N.~C.,  {Udry} S.,   {Mayor} M.,  2004,
  \mn@doi [\aap] {10.1051/0004-6361:20040384}, \href
  {http://adsabs.harvard.edu/abs/2004A%26A...426..695Z} {426, 695}

\bibitem[\protect\citeauthoryear{{da Silva} et~al.,}{{da Silva}
  et~al.}{2006}]{Sil2006}
{da Silva} L.,  et~al., 2006, \mn@doi [\aap] {10.1051/0004-6361:20065105},
  \href {http://adsabs.harvard.edu/abs/2006A%26A...458..609D} {458, 609}

\bibitem[\protect\citeauthoryear{{van Leeuwen}}{{van Leeuwen}}{2007}]{Van2007}
{van Leeuwen} F.,  2007, \mn@doi [\aap] {10.1051/0004-6361:20078357}, \href
  {http://cdsads.u-strasbg.fr/abs/2007A%26A...474..653V} {474, 653}

\makeatother
\end{thebibliography}

% -----------------------------------------------------------------------------

% Don't change these lines
\bsp	% typesetting comment
\label{lastpage}
\end{document}